\documentclass[aps,prc,groupedaddress,onecolumn,floatfix]{revtex4}
\usepackage{amsmath}
\usepackage{amssymb} 
\usepackage{amsfonts}
\usepackage{amscd}
\usepackage{bm}
\usepackage{graphicx}
\usepackage{color}
\usepackage{hyperref}
\hypersetup{colorlinks=true, citecolor=blue, urlcolor=blue}
\usepackage{empheq}

\def\beq{\begin{equation}}
\def\eeq{\end{equation}}

\begin{document}

\title{The relation between instant and light-front formulations of quantum
field theory} 

\author{W.~N.~Polyzou}
\email{polyzou@uiowa.edu}
\thanks{This work supported by the U.S. Department of Energy,
  Office of Science, Grant \#DE-SC0016457}
\affiliation{Department of Physics and Astronomy\\ The University of
Iowa\\ Iowa City, IA 52242, USA}

\date{\today}

\begin{abstract}

The scattering equivalence of quantum field theories formulated with
light-front and instant-form kinematic subgroups is established using
non-perturbative methods.  The difficulty with field theoretic
formulations of Dirac's forms of dynamics is that the free and
interacting unitary representations of the Poincar\'e group are
defined on inequivalent representations of the Hilbert space, which
means that the concept of kinematic transformations must be modified
on the Hilbert space of the field theory.  This work addresses this
problem by assuming the existence of a field theory with the expected
properties and constructs equivalent representations with instant and
front-form kinematic subgroups.  The underlying field theory is not
initially associated with an instant form or light-front form of the
dynamics.  In this construction the existence of a vacuum and
one-particle mass eigenstates is assumed and both the light-front and
instant-form representations are constructed to share the same vacuum
and one-particle states.  If there is spontaneous symmetry breaking
there will be a 0 mass particle in the mass spectrum (assuming no
Higgs mechanism).  The free field Fock space plays no role.  There is
no ``quantization'' of a classical theory.  The property that survives
from the perturbative approach is the notion of a kinematic subgroup,
which means kinematic Poincar\'e transformations can be trivially
implemented by acting on suitable basis vectors.  This
non-perturbative approach avoids dealing with issues that arise in
perturbative treatments where is it necessary to have a consistent
treatment of renormalization, rotational covariance, and the structure
of the light-front vacuum.  While addressing these issues in a
computational framework is the most important unanswered question for
applications, this work may provide some insight into the nature of
the expected resolution and identifies the origin of some of
differences between the perturbative and non-perturbative approaches.
  
\end{abstract}

\maketitle 

\section{Introduction}\label{sec1}

This paper discusses the relation between light-front and instant-form
formulations of quantum field theory.  The identification of different
forms of dynamics is due to Dirac \cite{Dirac:1949cp}.  In a
relativistic quantum theory the invariance of quantum probabilities in
different inertial coordinate systems requires that equivalent states
in different inertial coordinate systems are related by a unitary ray
representation of the subgroup of the Poincar\'e group continuously
connected to the identity \cite{Wigner:1939cj}.  The Poincar\'e Lie
algebra has 3 independent commutators involving rotationless boost and
translation generators that have the Hamiltonian on the right
\beq
[K^i,P^j]= i\delta_{ij} H.
\label{I:1}
\eeq
If $H=H_0+V$ for some interaction $V$ then the operators on the left
side of each commutator must also know about the interaction.  Dirac
identified three representations of the Poincar\'e Lie algebra with
the minimum number (3-4) of interaction-dependent Poincar\'e generators.  He
called these the instant, point and front-forms of the dynamics.  In
the instant form the generators of space translations and rotations
are free of interactions, in the point form the generators of
Lorentz transformations are free of interactions and in
the front form the generators of transformations that leave a hyperplane
tangent to the light cone invariant are free of interactions.

The equivalence of these different representations of relativistic
quantum mechanics was settled by Sokolov and Shatnyi
\cite{Sokolov:1977im}\cite{Keister:1996bd}.  In quantum field theory
the problem is more complicated because the free and interacting
dynamics are formulated on different inequivalent representations of
the Hilbert space \cite{Haag:1955ev}, so the decomposition of the
Hamiltonian into the sum of a free Hamiltonian plus interaction is not
defined on the Hilbert space of the field theory.  Such a
decomposition makes
sense in perturbative quantum field theory with cutoffs, so the notion
of instant- and light-front formulations
\cite{Chang1069}\cite{Soper}\cite{Chang:1972xt}\cite{Yan:1973.2}\cite{Yan:1973.3}\cite{Yan:1973.4} of
quantum field theory make sense perturbatively.  The price paid is
that as the cutoffs are removed the theory has infinities that have to
be renormalized.  The relation between the different forms of dynamics
depends on a consistent treatment of the renormalization.  The
light-front formulation of quantum field theory is of particular interest for
applications.  The most appealing property of the theory is the
apparent triviality of the light-front vacuum, which reduces the
solution of the field theory to linear algebra on a Hilbert space,
like non-relativistic quantum mechanics.  In addition to the
computational challenges of implementing this program in a theory with
an infinite number of degrees of freedom, there are a number of
puzzles that appear in comparing the two approaches.  These include:

\begin{itemize}
\item[$\bullet$] Is the light-front vacuum the same as the Fock vacuum?

\item[$\bullet$] How to understand $P^+=0$ (zero mode) singularities?

\item[$\bullet$] How to understand spontaneous symmetry breaking in a light-front
  dynamics?
  
\item[$\bullet$] How to renormalize the theory consistent with rotational covariance.

\item[$\bullet$] What is the relation between light-front and canonical quantization of
  a quantum field theory?

\item[$\bullet$] Are both approaches equivalent?  
\end{itemize}
While there is a general consensus that the two formulations are
equivalent, the answers to the above questions are not as clean as
desired.  These issues have been discussed extensively in the
literature
\cite{Chang1069}
\cite{Soper}
\cite{Leutwyler:1970wn}
\cite{Rohrlich:1971zz}
\cite{Chang:1972xt}
\cite{Yan:1973.2}
\cite{Yan:1973.3}
\cite{Yan:1973.4}
\cite{Schlieder:1972qr}
\cite{Maskawa:1976}
\cite{Nakanishi:1976yx}
\cite{Nakanishi:1977}
\cite{Leutwyler:1977vy}
\cite{Coester:1992}
\cite{Coester:1993fg}
\cite{Bylev:1996}
\cite{Brodsky:1998}
\cite{Yamawaki:1998}
\cite{choi:1998}
\cite{Tsujimaru:1997jt}
\cite{Lenz:2000}
\cite{Heinzl:2001}
\cite{Burkhardt:2002}
\cite{Heinzl:2003}
\cite{Werner:2006}
\cite{Bakker:2011zza}
\cite{Choi:2011xm}
\cite{Choi:2013ira}
\cite{Beane:2013oia}
\cite{sofia:2014}
\cite{Herrmann:2015dqa}
\cite{Ji:2017vfu}
\cite{Collins:2018aqt}
\cite{Mannheim:2020rod}.

Most of these discussions are based on perturbation theory; in
particular on the assumption that the light-front dynamics can be
formulated on the Fock space of a free field theory.  The continued
interest in these questions is because a compelling resolution of
these questions is obscured by the need to perform a non-perturbative
renormalization of the theory.  In this work these complications are
avoided by assuming the existence of the theory with the expected
properties that are normally assumed in axiomatic formulations of
quantum field theory
\cite{Wightman:1980}\cite{Haag:1963dh}\cite{Osterwalder:1973dx}\cite{Osterwalder:1974tc}.
One departure from both of Dirac's Hamiltonian approaches is that in
axiomatic formulations of quantum field theory states are constructed
by applying fields integrated against Schwartz test functions
\cite{Gelfand} in four space-time variables to the vacuum.  These
states can be evolved, but the initial data is not on a sharp time
slice or light front.  The existence of an asymptotically complete
scattering theory is also assumed, based on the Haag-Ruelle
formulation of scattering theory
\cite{Haag:1958vt}\cite{Brenig:1959}\cite{Ruelle:1962}\cite{jost:1966}.
Haag-Ruelle scattering has the advantage that it can be formulated in
terms of wave operators defined by strong limits.  The strong limits
justify using many of the methods that are used in non-relativistic
scattering theory and they are used in the construction that follows.
The difference with the LSZ formulation of scattering is that the
interpolating field operators must create one-particle states rather
than just create states that have some overlap with one-particle
states.  The advantage of the Haag-Ruelle formulation of scattering is
that the weak limits that are used to define the scattering asymptotic
conditions in the LSZ formulation of scattering can be replaced by
strong limits.

A two-Hilbert space representation, which does not require the
existence of a free dynamics on the Hilbert space of the theory, is
used to formulate the scattering theory.  This avoids any concerns
about Haag's theorem \cite{Wightman:1980}\cite{Haag:1955ev}, which
implies that the interacting and free field theories are formulated on
inequivalent Hilbert-space representations of the field algebra.  The
second Hilbert space, which is called the asymptotic Hilbert space, is
the direct sum of tensor products of single-particle Hilbert spaces
with physical masses.  This representation is natural in
non-relativistic many-body scattering.  For example, in lead-lead
scattering the lead nuclei are treated as elementary particles in the
asymptotic Hilbert space.  Products of the lead state vectors, $\vert
L_1, \mathbf{p}_1,\mu_1 \rangle \times \vert L_2,\mathbf{p}_2,\mu_2
\rangle$, define a mapping from the asymptotic Hilbert space of
suitably symmetrized square integrable functions, $f(
\mathbf{p}_1,\mu_1,\mathbf{p}_2,\mu_2)$, to the Hilbert space of the
theory.  In the field theory case the lead wave functions are replaced
by products of operators, $A_L^{\dagger}(\mathbf{p},\mu)$, that create
point spectrum lead mass eigenstates out of the vacuum, applied to the
vacuum \cite{simon} \cite{araki-bk}\cite{baumgartl:1983}.  In the
field theory case the operators $A_L^{\dagger}(\mathbf{p},\mu)$ are
not creation operators.  They do not commute or anticommute
with each other and do
not satisfy canonical commutation relations.  Inner products of states
constructed by applying two or more of these operators to the vacuum
can have intermediate states with different particle content.

In the simplest case the asymptotic Hilbert space is a Fock space with
physical particle masses.  The mapping from the second Hilbert space
to the Hilbert space of the field theory puts in the internal
structure of the physical particles, including the effects of
self-interactions, when the particles are asymptotically separated.

There is a natural unitary representation of the Poincar\'e group on
the asymptotic Hilbert space, which treats the particles as free
particles with their observed masses.  There is sufficient freedom to
choose the mappings from the asymptotic Hilbert space to the Hilbert
space of the field theory so the unitary representation of the instant
or light-front kinematic subgroup of the Poincar\'e group on the
asymptotic space is mapped to the dynamical unitary representation of
the corresponding kinematic subgroup on the field-theory Hilbert space without
changing the scattering operator.


A theorem due to Ekstein \cite{Ekstein:1960} gives necessary and
sufficient conditions for unitary transformations on a Hilbert space
to preserve the scattering operator.  This result assumes the
scattering is formulated in terms of asymptotically complete wave operators defined as strong
limits.  In the two-Hilbert space formulation Ekstein's condition is a
constraint on the unitary transformation relating two different
Hamiltonians that does not require modifying the Hamiltonian in the
asymptotic Hilbert space.  Roughly speaking it means that the unitary
operator is a short-range perturbation of the identity in a manner that Ekstein
makes precise (see eq. (\ref{S1:33})).  This freedom can be exploited
to construct the most general class of S-matrix preserving unitary
transformations that are invariant under a kinematic subgroup of the
dynamical representation of the Poincar\'e group.  The proof uses the
unitary of the wave operators which holds provided the vacuum and
one-body channels are included in the asymptotic Hilbert space.
Unitary transformations must leave the one-particle
masses and the mass gap unchanged.  In this work these transformations
are also constructed to leave the vacuum and one-particle states unchanged.

When these unitary transformations are applied to the unitary
representation of the Poincar\'e group on the physical Hilbert space,
they result in two new scattering-equivalent representations with
light-front and instant-form kinematic subgroups.  These
representations have the property that the injection operators map the
kinematic subgroup on the asymptotic Hilbert space to the
corresponding subgroup of the unitary representation of the Poincar\'e
group on the physical Hilbert space.  Both representations are
equivalent to the original ``axiomatic'' formulation of the theory
in the sense that they are related by $S$-matrix preserving unitary
transformations.

One thing that is missing in this construction is the existence of
a non-interacting representation of the Poincar\'e group on the
Hilbert space of the field theory.  Haag's theorem implies that free
fields, even with experimental masses, are defined on an inequivalent
Hilbert space representation of the field algebra, so there is no
corresponding local free field theory on the Hilbert space of the
field theory. In the event that the mapping from the asymptotic space
to the field theory Hilbert space has dense range, and the unitarized
form of this mapping satisfies Ekstein's asymptotic condition,
eq. (\ref{S1:33}), then it can be used to map the unitary
representation of the Poincar\'e group on the asymptotic Hilbert space
to a unitary representation of the Poincar\'e group on the field
theory Hilbert that plays the role of a free-particle dynamics.  Since
the two-Hilbert space injection operators are not unique and are not
local operators, this will not be a local field theory.

The results can be summarized as follows. The construction assumes a
unitary representation of the Poincar\'e group on the Hilbert space of
the field theory and constructs scattering equivalent representations
with light-front and instant-form kinematic symmetries.  The
representations are related by a unitary transformation that satisfies
an asymptotic condition that preserves the scattering matrix elements.
Because the one-body states are eigenstates of the mass operator,
there is no mass renormalization.  While the form of the interactions
is representation dependent, the
absence of mass renormalization suggests the that simplest types of
interactions that change particle number are short-range
$2\leftrightarrow 3$-particle interactions rather than
$1\leftrightarrow 2$ vertex interactions.

This paper has five sections and two appendices.  Section two
summarizes the assumptions used in this work and discusses a
two-Hilbert space formulation of Haag-Ruelle scattering that has
kinematically invariant injection operators that map an asymptotic
many-particle Hilbert space to the Hilbert space of the field theory.
Haag-Ruelle scattering is the natural generalization of the usual
formulation of time-dependent scattering theory.  The construction of
equivalent instant and light-front representations of the field theory
is given in section 3.  Section 4 has a brief discussion of
spontaneous symmetry breaking.  The construction is compared to the
corresponding construction in relativistic formulations of quantum
mechanics of a finite number of degrees of freedom in section 5.
Section 6 has a summary of the results and conclusions. There are two
appendices.  The first one contains a discussion of Ekstein's
treatment of scattering equivalences \cite{Ekstein:1960} that is used
in section 3.  The second appendix discusses the construction of
Haag-Ruelle injection operators with the kinematic symmetry properties
that are used in sections two and three.

\section{General considerations}

This paper examines the relation between light-front and instant-form
formulations of quantum field theory from a more abstract perspective
that does not assume that the Hamiltonian can be decomposed as the sum of 
free and interacting operators acting on one representation of the
Hilbert space.  The starting assumptions are typical assumptions about
abstract properties of a quantum field theory.  These include:

\begin{itemize}
  
\item[$\bullet$] The Hilbert space of the field theory is generated by
  applying bounded functions, $F$, of field operators smeared
  over test functions to a
  unique vacuum vector, denoted by $\vert 0 \rangle$.  The smearing is
  assumed to be over Schwartz test functions \cite{Gelfand} in four space-time
  dimensions.  It is not assumed that fields
  restricted to a light-front or fixed-time manifold make sense.  A
  dense set of vectors, $\vert \psi \rangle$, can be taken to have the
  form:

\beq
\vert \psi \rangle = F \vert 0 \rangle
\qquad 
F=\sum_{n=1}^N c_n e^{i\phi_n (f_n)}
\label{S1:1}
\eeq
\item[]where the $c_n$ are complex coefficients, the $f_n$ are Schwartz
functions, $\phi_n$ are field operators and $N$ is finite.

\item[$\bullet$] There is a unitary representation of the Poincar\'e
  group, $U(\Lambda,a)$, on this representation of the Hilbert space
  which is given by covariance.  In the spinless case the action of
  $U(\Lambda ,a)$ on states of the form (\ref{S1:1}) is
\beq
U(\Lambda ,a) \vert \psi \rangle = \sum_c c_n e^{i\phi (f_n')} \vert 0 \rangle
\qquad
f_n'(x) = f_n(\Lambda^{-1}(x-a)).
\label{S1:2}
\eeq

\item[$\bullet$] The vacuum, $\vert 0 \rangle$, is Poincar\'e invariant and normalized to unity:
\beq
U(\Lambda ,a)\vert 0 \rangle  = \vert 0 \rangle \qquad \langle 0 \vert 0 \rangle =1 .
\label{S1:3}
\eeq

\item[$\bullet$] The unitary representation of the Poincar\'e group acts like the
identity on the vacuum subspace and can be decomposed into a direct integral
of positive-mass positive-energy irreducible representations of the
Poincar\'e group \cite{Wigner:1939cj}\cite{Bargmann:1946me} on the orthogonal complement of the
vacuum subspace.  The mass Casimir operator is assumed to have point-spectrum
eigenstates representing particles of the theory.

\item[$\bullet$] There is a complete set of asymptotically
  complete Haag-Ruelle \cite{Haag:1958vt}\cite{Brenig:1959}\cite{Ruelle:1962}\cite{jost:1966}
scattering states.

\end{itemize}

A characteristic element of Dirac's forms of dynamics is the notion of
a {\it kinematic subgroup}.  Kinematic subgroups leave a manifold in
Minkowski space, that classically intersects every (massive)
particle's world line once, invariant.  In an instant-form dynamics
the invariant manifold is a fixed-time plane, which is preserved under
the group generated by rotations and space translations.  In a
light-front dynamics the manifold is a hyperplane that is tangent to the
light cone.  This is invariant under a seven-parameter subgroup of the
Poincar\'e group.  These manifolds are 
prominent in Dirac's work; in this work the focus will be on the subgroups of
the Poincar\'e group that leave these manifolds invariant.  These
subgroups will be referred to as kinematic subgroups.

For quantum mechanical systems of a finite number of degrees of
freedom kinematical and dynamical unitary representations of the
Poincar\'e group exist in the same representation of the Hilbert
space.  In the field theoretic case the representation of the Hilbert
space depends on the dynamics.  While free fields with physical masses
have the same mass spectrum, they act on an inequivalent
Hilbert space representation of the field algebra by Haag's theorem,
so {\it there is no free dynamics} on the Hilbert space of the
interacting field theory.  This is one of the
distinctions with the perturbative approach, which attempts to define
the interacting theory as a perturbation of a free field theory.  The
incompatibility of these formulations leads to the infinities that
complicate the analysis of the relation between light-front and
instant-form treatments of field theory.

The first step in the abstract  formulation is to define what is meant
by  the  kinematic  subgroup  in  the  absence  of  a  non-interacting
representation of the Poincar\'e group.  The unitary representation of
the Poincar\'e group has a set of 10 infinitesimal generators that are
self-adjoint  operators  on the  Hilbert  space  of the  field  theory
satisfying the Poincar\'e Lie algebra.   This follows because each one
is  the  generator of  a  unitary  one-parameter group.   The  unitary
representation of the Poincar\'e group can be decomposed into a direct
integral  of  irreducible representations.   In  this  work a  typical
spectral  condition is assumed,  where the  representations that
appear  in  the  direct  integral  are  assumed  to  be  positive-mass
positive-energy  representations,  and  the  identity  on  the  vacuum
subspace. 

The direct integral of irreducible representations of the Poincar\'e
group results in a direct integral decomposition of the Hilbert space into
irreducible subspaces or slices.  The next step is to choose a basis
on each irreducible subspace.  Two choices will be considered in this
work.  The first basis consists of simultaneous eigenstates of the
mass, spin, linear momentum, and the projection of the canonical spin
on the $z$ axis.  These are all functions of the infinitesimal
generators of the unitary representation of the Poincar\'e group.  There
will also be Poincar\'e invariant degeneracy parameters, which can be
taken as discrete quantum numbers, since any continuous parameters can
be replaced by a basis of square integrable functions of the
continuous degeneracy parameters.  The other basis on the same
irreducible subspace consists of simultaneous eigenstates of the mass,
spin, the three light-front components of the four momentum, and the
projection of the light-front spin on the $z$ axis.  These are also
functions of the infinitesimal generators with the same degeneracy
parameters.

These bases can be formally expressed as
\beq
\vert 0 \rangle \cup \{
\vert (m,s,d_n) \mathbf{p},\mu_c \rangle \} \qquad \mbox{and} \qquad
\vert 0 \rangle \cup \{ \vert (m,s,d_n) \tilde{\mathbf{p}},\mu_F
\rangle \}
\label{S1:4}
\eeq
where the light-front components of the four momentum are defined by
\beq
\tilde{\mathbf{p}} =
(\mathbf{p}\cdot \hat{\mathbf{x}}, \mathbf{p}\cdot \hat{\mathbf{y}}, p^0+\mathbf{p}\cdot \hat{\mathbf{z}}) :=
(\mathbf{p}_{\perp},p^+).
\label{S1:5}
\eeq
The variables (\ref{S1:5}) are eigenvalues of the generators of
translations tangent to the light-front hyperplane.  The first basis
in (\ref{S1:4}) will be referred to as the ``instant-form'' basis and
the second basis will be referred to as the ``light-front'' basis.

These basis states are related by a unitary change of basis.  The
change of basis is block diagonal in the direct integral.  The
different basis vectors are related by
\beq
\vert (m,s,d_n) \mathbf{p},\mu_I \rangle =
\sum_{\mu_f=-s}^s  \vert (m,s,d_n) \tilde{\mathbf{p}}(p),\mu_F \rangle
\sqrt{p^+ \over \omega_m(\mathbf{p})}
D^s_{\mu_F \mu_I}[B_F^{-1}(p/m) B_I(p/m)]
\label{S1:6}
\eeq
which assumes that both bases have delta function normalizations:
\beq
\langle (m,s,d_n) \mathbf{p},\mu_I \vert (m',s',d_k)
\mathbf{p}',\mu'_I \rangle =
\delta (\mathbf{p}-\mathbf{p}')\delta_{ss'}\delta_{\mu_I\mu_I'} \delta_{nk}
\delta [m-m']
\label{S1:7}
\eeq
and
\beq
\langle (m,s,d_n) \tilde{\mathbf{p}},\mu_F \vert (m',s',d_k)
\tilde{\mathbf{p}}',\mu'_F \rangle =\delta (p^+-p^{+\prime})
\delta (\mathbf{p}_{\perp}-\mathbf{p}_{\perp}')\delta_{ss'}\delta_{\mu_F\mu_F'} \delta_{nk}
\delta [m-m']
\label{S1:8}
\eeq
where $\delta [m-m']$ is either a Dirac or Kronecker delta function
depending on whether $m$ is in the point or continuous spectrum of the
mass operator.  $B_I(p/m)$ and $B_F(p/m)$ are $SL(2,C)$ matrices
representing a rotationless boost and a light-front preserving boost
from $(1,0,0,0)$ to $p/m$.  The combination $B_F^{-1}(p/m) B_I(p/m)$
is a $SU(2)$ representation of a Melosh rotation \cite{Melosh:1974cu}
that changes the light-front spin to the canonical spin.
$
D^s_{\mu_F \mu_I}[R] =
\langle s,\mu_F \vert U(R) \vert s, \mu_I \rangle$ is the $2s+1$ dimensional
unitary representation of $SU(2)$ and 
$\omega_m(\mathbf{p})=\sqrt{\mathbf{p}^2 +m^2}$.

There are similar transformations relating light-front bases
corresponding to different orientations of the light front
(i.e. replacing $\hat{\mathbf{z}}$ by some other unit vector
$\hat{\mathbf{n}}$).  Like (\ref{S1:6}) they involve a variable
change, the square root of a Jacobian and a momentum-dependent
rotation matrix.

The action of $U(\Lambda ,a)$ on each of these irreducible basis
states is a consequence of the transformation properties of the
Poincar\'e generators and the basis choice.
For the ``instant-form'' basis
\beq
U(\Lambda ,a) \vert 0 \rangle = \vert 0 \rangle ,
\label{S1:10}
\eeq
\beq
U(\Lambda ,a) \vert (m,s,d_n) {\mathbf{p}},\mu \rangle
=
e^{ia \cdot \Lambda p}  \sum_{\mu'=-s}^s
\vert (m,s,d_n) {\pmb{\Lambda}{p}},\mu' \rangle
\sqrt{{ \omega_m (\pmb{\Lambda}p) \over \omega_m (\mathbf{p})}}
D^s_{\mu'\mu}[B_I^{-1}(\Lambda ( p/m)) \Lambda B_I(p/m)]
\label{S1:11}
\eeq
and for the ''light-front'' basis,
\beq
U(\Lambda ,a) \vert 0 \rangle = \vert 0 \rangle ,
\label{S1:12}
\eeq
\beq
U(\Lambda ,a) \vert (m,s,d_n) {\tilde{\mathbf{p}}},\mu \rangle
=
e^{ia \cdot \Lambda p} \sum_{\mu'=-s}^s
\vert (m,s,d_n) {\tilde{\pmb{\Lambda}}{p}},\mu' \rangle
\sqrt{{(\Lambda p)^+ \over p^+}}
D^s_{\mu'\mu}[B_F^{-1}(\Lambda (p/m)) \Lambda B_F(p/m)].
\label{S1:13}
\eeq
These equations define the dynamical
unitary representation of the Poincar\'e group on these two
irreducible bases.

The connection of these bases with kinematic subgroups is that the
coefficients of the basis functions on the right hand side of
equations (\ref{S1:11}) and (\ref{S1:13}) do not depend on the mass
eigenvalue $m$ when $(\Lambda,a)$ is an element of the kinematic
subgroup.  For the basis (\ref{S1:11}) the kinematic subgroup is
generated by rotations and spatial translations while for the basis
(\ref{S1:13}) the kinematic subgroup is the subgroup that leaves the
light front hyperplane $x^+=x^0 + \mathbf{x} \cdot \hat{\mathbf{z}}=0$
invariant.

The next step is to discuss the formulation of scattering theory.  A
two-Hilbert space representation \cite{Coester:1982vt}\cite{baumgartl:1983}, will be used for this purpose.  The
starting point is the assumption that the mass operator,
$M=\sqrt{-p^2}$ has one-particle eigenstates.  In this
non-perturbative context one-particle means that the mass operator, $M
= \sqrt{-p^2}$, has a non-empty point spectrum. The discrete mass
eigenvalues are assumed to be strictly positive.
There is no distinction between elementary and composite particles.

Normalizable one-particle states in the ``instant form'' basis
can be constructed by integrating the irreducible basis functions with a
wave packet:
\beq
\vert \psi_{Ig} \rangle := \int d\mathbf{p} \sum_{\mu=-s}^s
\vert  (m,s,d_n) \mathbf{p},\mu \rangle g (\mathbf{p},\mu)
\label{S1:14}
\eeq
where $m$ is a discrete mass eigenvalue and $g (\mathbf{p},\mu)$
is a square integrable wave packet.
The corresponding construction in
the ``light-front'' basis has the form
\beq
\vert \psi_{Fg} \rangle := \int d^2\mathbf{p}_{\perp} \int_0^\infty dp^+
\sum_{\mu=-s}^s
\vert  (m,s,d_n) \tilde{\mathbf{p}},\mu \rangle \tilde{g} (\tilde{\mathbf{p}},\mu)
\label{S1:15}
\eeq
where in the light front case $\tilde{\mathbf{p}}$ should vanish as
$p^+ \to 0$ in a manner that results in a square integrable function
of $\mathbf{p}$ under the variable change (\ref{S1:5}) $\tilde{\mathbf{p}}\to
{\mathbf{p}}$.

These normalizable vectors are linear in the functions
${g} ({\mathbf{p}},\mu)$ and $ \tilde{g} (\tilde{\mathbf{p}},\mu)$.
They
can be represented by elements of the field algebra applied to the
vacuum. This means that they can be expressed as
\beq
\vert \psi_{Ig} \rangle  = \sum_{\mu=-s}^s \int A^{\dagger}_{m,s,d_n}({p},\mu)\vert 0 \rangle d\mathbf{p}
g (\mathbf{p},\mu) = A^{\dagger}(g)\vert 0 \rangle
\label{S1:16}
\eeq
and
\beq
\vert \psi_{Fg} \rangle  = \sum_{\mu=-s}^s \int \tilde{A}^{\dagger} _{m,s,d_n}(\tilde{{p}},\mu)\vert 0 \rangle dp^+ d^2\mathbf{p}_{\perp}
\tilde{g} (\tilde{\mathbf{p}},\mu) = \tilde{A}^{\dagger}(\tilde{g})\vert 0 \rangle .
\label{S1:17}
\eeq
These states will be equal if $g ({\mathbf{p}},\mu) =
\langle (m,s,d_n) \mathbf{p},\mu_I \vert g \rangle $ and
$\tilde{g} (\tilde{\mathbf{p}},\mu)
= \langle (m,s,d_n) \tilde{\mathbf{p}},\mu_F \vert g \rangle $
are related by the change of basis (\ref{S1:6}).  

The operators $A^{\dagger}_{m,s,d_n}({p},\mu)$ and
$\tilde{A}^{\dagger}_{m,s,d_n}(\tilde{{p}},\mu)$ are functions of the
fields that create particles of mass $m$, spin $s$, momentum
$\mathbf{p}$ and magnetic quantum number $\mu$ {\it out of the vacuum}.
The construction of these operators starting from operators that couple
the vacuum to the one-particle states of the theory is discussed in
appendix \ref{inj}.

Since $A^{\dagger}_{m,s,d_n}({p},\mu)$ and
$\tilde{A}^{\dagger}_{m,s,d_n}(\tilde{{p}},\mu)$ are in the field
algebra (after smearing with test functions), they can be multiplied.

The following quantity
\beq
A^{\dagger}_{m_1,s_1,d_{n_1}}(\mathbf{p_1},\mu_1) \cdots A^{\dagger}_{m_k,s_k,d_{n_k}}
(\mathbf{p_k},\mu_k) \vert 0 \rangle 
\label{S1:18}
\eeq
can be considered as a mapping from a Hilbert space of square integrable
suitably symmetrized functions of $\mathbf{p}_1, \mu_1 \cdots \mathbf{p}_k,\mu_k$ to the
Hilbert space of the field theory.  Note that while
$A^{\dagger}(g)\vert 0 \rangle = \tilde{A}^{\dagger}
(\tilde{g})\vert 0 \rangle$ for $g$ and $\tilde{g}$
related by (\ref{S1:6}), because both operators create the same
single-particle states out of the vacuum,
this identity is not true for products of these
operators applied to the vacuum.
From equation (\ref{A2:17}) in
appendix \ref{inj} it follows that this mapping has the following
properties
\[
\mathbf{P} A^{\dagger}_{m_1,s_1,d_{n_1}}(\mathbf{p}_1,\mu_1) \cdots
A^{\dagger}_{m_N,s_N,d_{n_N}}(\mathbf{p}_N,\mu_N)
\vert 0 \rangle =
\]
\beq
\sum_{n=1}^N\mathbf{p}_n
A^{\dagger}_{m_1,s_1,d_{n_1}}(\mathbf{p}_1,\mu_1) \cdots
A^{\dagger}_{m_N,s_N,d_{n_N}}(\mathbf{p}_N,\mu_N)
\vert 0 \rangle
\label{S1:19}
\eeq
and
\[
U(R,0 )A^{\dagger}_{m_1,s_1,d_{n_1}}(\mathbf{p}_1,\mu_1) \cdots
A^{\dagger}_{m_N,s_N,d_{n_N}}(\mathbf{p}_1,\mu_N)
\vert 0 \rangle
 =
\]
\beq
\sum
A^{\dagger}_{m_1,s_1,d_{n_1}}(R\mathbf{p}_1,\nu_1) \cdots
A^{\dagger}_{m_N,s_N,d_{n_N}}(R\mathbf{p}_N,\nu_N)
\vert 0 \rangle 
\vert 0 \rangle
\prod_{n=1}^N D^{s_n}_{\nu_n \mu_n}(R) .
\label{S1:20}
\eeq
For the mapping in the
``light-front'' basis the corresponding relations are 
\[
\tilde{\mathbf{P}} \tilde{A}^{\dagger}_{m_1,s_1,d_{n_1}}(\tilde{\mathbf{p}}_1,\mu_1) \cdots \tilde{A}^{\dagger}_{m_N,s_N,d_{n_N}}(\tilde{\mathbf{p}}_N,\mu_N) \vert 0 \rangle =
\]
\beq
\sum_{n=1}^N\tilde{\mathbf{p}}_n \tilde{A}^{\dagger}_{m_1,s_1,d_{n_1}}(\tilde{\mathbf{p}}_1,\mu_1) \cdots \tilde{A}^{\dagger}_{m_N,s_N,d_{n_N}}(\tilde{\mathbf{p}}_N,\mu_N) \vert 0 \rangle 
\label{S1:21}
\eeq
\[
U(B_f,0) \tilde{A}^{\dagger}_{m_1,s_1,d_{n_1}}(\tilde{\mathbf{p}}_1,\mu_1) \cdots
\tilde{A}^{\dagger}_{m_N,s_N,d_{n_N}}(\tilde{\mathbf{p}}_N,\mu_N) \vert 0 \rangle =
\]
\beq
\tilde{A}^{\dagger}_{m_1,s_1,d_{n_1}}(\tilde{\mathbf{B}}_f p_1,\mu_1) \cdots
\tilde{A}^{\dagger}_{m_N,s_N,d_{n_N}}(\tilde{\mathbf{B}}_f p_N,\mu_N) \vert 0 \rangle
\prod_{n=1}^N {\sqrt {B_f(p_n)^+\over p_n^+}}.
\label{S1:22}
\eeq
\[
U(R_z(\phi),0) \tilde{A}^{\dagger}_{m_1,s_1,d_{n_1}}(\tilde{\mathbf{p}}_1,\mu_1) \cdots
\tilde{A}^{\dagger}_{m_N,s_N,d_{n_N}}(\tilde{\mathbf{p}}_N,\mu_N) \vert 0 \rangle =
\]
\beq
\tilde{A}^{\dagger}_{m_1,s_1,d_{n_1}}(R_z(\phi)\tilde{\mathbf{p}}_1,\mu_1) \cdots
\tilde{A}^{\dagger}_{m_N,s_N,d_{n_N}}(R(\phi) \tilde{\mathbf{p}}_N,\mu_N) \vert 0 \rangle
\prod_{n=1}^N e^{i \mu_n \phi}.
\label{S1:23}
\eeq

If the Fourier transform of these states are integrated against wave
packets that are localized in regions separated by large space-like
separations and the field theory satisfies cluster properties
(normally a consequence of uniqueness of the vacuum) then the
resulting vectors look like states of $N$
asymptotically separated particles when they are used in
inner products.  This property is used
\cite{jost:1966}\cite{araki-bk} to prove the existence of the strong
limits that define wave operators in Haag-Ruelle scattering.
The interpretation of the operators
${A}^{\dagger}_{m,s,d_{n}}$ and  $\tilde{A}^{\dagger}_{m,s,d_{n}}$
is that they {\it asymptotically} behave like creation operators .

A scattering channel $\alpha$ is associated with a finite collection
of particles asymptotically.  It could correspond to the state of a
target and an incoming projectile or a collection of particles that are
detected in a scattering experiment. The set of all scattering
channels of the theory is denoted by ${\cal A}$.  In what follows both
the vacuum and one-particle states are included in the collection of
channels, ${\cal A}$.

Channel injection operators are defined as
products of the operators (\ref{S1:16}) or (\ref{S1:17}),
applied to the vacuum,  one for each
particle in a scattering channel $\alpha$.  In the canonical
basis they are 
\beq
\Phi_{\alpha} ({\mathbf{p}_1},\mu_1 \cdots {\mathbf{p}_N},\mu_N):=
\prod_{k \in \alpha}  A^{\dagger}_{m_k,s_k,d_{n_k}}({\mathbf{p}_k},\mu_k)\vert 0 \rangle
\label{S1:24}
\eeq
and in the light-front basis they are
\beq
\tilde{\Phi}_{\alpha} (\tilde{\mathbf{p}}_1,\mu_1 \cdots \tilde{\mathbf{p}}_N,\mu_N):=
\prod_{k \in \alpha}
\tilde{A}^{\dagger}_{m_k,s_k,d_{n_k}}(\tilde{\mathbf{p}}_k,\mu_k)\vert0 \rangle . 
\label{S1:25}
\eeq
The asymptotic channel Hilbert space ${\cal H}_{\alpha}$ is the space
of square integrable functions of the variables $\mathbf{p}_k,\mu_k$
or $\tilde{\mathbf{p}}_k,\mu_k$ for $k \in \alpha$.  It is interpreted
as a space of $N$ particles of mass $m_k$ and spin $s_k$.  If any of the
particles in the channel $\alpha$ are identical then the functions representing
identical Bosons should be symmetrized and those representing
identical Fermions should anti-symmetrized.  
The order of the product in (\ref{S1:25}) does not matter for
suitably symmetrized states in the
asymptotic Hilbert space.
The channel injection
operators (\ref{S1:24}) and (\ref{S1:25}) are interpreted as mappings
from the $k$-particle channel subspace of the asymptotic
Hilbert space ${\cal H}_{\alpha}$ to the Hilbert
space ${\cal H}$ of the field theory.

The asymptotic unitary representation of the Poincar\'e group
on ${\cal H}_{\alpha}$ is defined by treating ${\cal H}_{\alpha}$ as a space of
$N$ mutually non-interacting particles of mass $m_k$ and spin $s_k$
for $k \in \alpha$.  This representation is defined by the tensor
product of one-particle irreducible representations in terms of
``instant-form'' variables
\beq
U_\alpha(\Lambda ,a)\vert \mathbf{p}_1, \mu_1, \cdots
\mathbf{p}_k, \mu_k \rangle  :=\sum
e^{ia \cdot \Lambda (\sum_n p_n)}
\vert \pmb{\Lambda}p_1, \nu_1 \cdots \pmb{\Lambda}p_k, \nu_k \rangle
\prod_{n=1}^k
\sqrt{{ \omega_{m_n} (\pmb{\Lambda}p_n) \over \omega_{m_n} (\mathbf{p}_n)}}
D^s_{\nu_n\mu_n}[B_c^{-1}(\Lambda p_n/m_n) \Lambda B_c(p_n/m_n)].
\label{S1:26}
\eeq
The corresponding expression in terms of ``light-front'' variables is
\beq
U_\alpha(\Lambda ,a)\vert \tilde{\mathbf{p}}_1, \mu_1, \cdots
\tilde{\mathbf{p}}_k, \mu_k \rangle  :=\sum
e^{ia \cdot \Lambda (\sum_n p_n)}
\vert \tilde{\pmb{\Lambda}}p_1, \nu_1 \cdots \tilde{\pmb{\Lambda}}p_k, \nu_k \rangle
\prod_{n=1}^k
\sqrt{{(\tilde{\Lambda}p_n)^+ \over {p}^+_n}}
D^s_{\nu_n\mu_n}[B_f^{-1}(\Lambda p_n/m_n) \Lambda B_f(p_n/m_n)].
\label{S1:27}
\eeq
The states in (\ref{S1:26}) and (\ref{S1:27}) are basis vectors in
${\cal H}_{\alpha}$, not ${\cal H}$.  Since the Hilbert space vectors
in the range of $\Phi_{\alpha}$ or $\tilde{\Phi}_{\alpha}$ are not
necessarily N-particle states it follows that in general
\beq
U(\Lambda,a)\Phi_{\alpha} \not= \Phi_{\alpha}U_{\alpha}(\Lambda,a).
\qquad
U(\Lambda,a)\tilde{\Phi}_{\alpha} \not= \tilde{\Phi}_{\alpha}U_{\alpha}(\Lambda,a).
\label{S1:28}
\eeq
This is because 
(\ref{A2:6}) does not hold for $P^0$ due to the
$p^0$ integrals in (\ref{A2:9}) and 
(\ref{A2:6}) does not hold for $P^-$ due to the
$p^-$ integrals on (\ref{A2:9}) (see appendix \ref{inj}).

When $\Lambda_K$ and $a_K$ are elements of the
instant-form or light-front kinematic subgroup,
it follows from (\ref{S1:19}-\ref{S1:23}) that 
\beq
U(\Lambda_K,a_K)\Phi_{\alpha} = \Phi_{\alpha}U_{\alpha}(\Lambda_K,a_K)
\qquad
U(\Lambda_K,a_K)\tilde{\Phi}_{\alpha} = \tilde{\Phi}_{\alpha}U_{\alpha}(\Lambda_K,a_K)
\label{S1:29}
\eeq
because the coefficients of the kinematic transformations are mass
independent.  The operators $\Phi_{\alpha}$ and
$\tilde{\Phi}_{\alpha}$ that map the channel $\alpha$ Hilbert space
${\cal H}_{\alpha}$ into the field theory Hilbert space ${\cal H}$ are
Haag-Ruelle \cite{Haag:1958vt}\cite{Brenig:1959}\cite{Ruelle:1962}
injection operators.  The only difference with the
standard choices of Haag Ruelle injection operators is that
$\Phi_{\alpha}$ and
$\tilde{\Phi}_{\alpha}$ 
are designed
to preserve the kinematic subgroup (\ref{S1:29}).

In the two-Hilbert space 
formulation \cite{simon}\cite{araki-bk} channel wave operators $\Omega_{\alpha \pm} =
\Omega_{\alpha \pm} (H,\Phi_{\alpha},H_{\alpha})$ are defined by the
strong limits
\beq
\lim_{t \to \pm\infty}
\Vert (\Omega_{\alpha \pm} (H,\Phi_{\alpha},H_{\alpha}) -
e^{iHt} \Phi_{\alpha} e^{-i H_\alpha t}) \vert \psi \rangle \Vert =0
\label{S1:30a}
\eeq
or
\beq
\lim_{t \to \pm\infty}
\Vert (\Omega_{\alpha \pm} (H,\tilde{\Phi}_{\alpha},H_{\alpha}) -
e^{iHt} \tilde{\Phi}_{\alpha} e^{-i H_\alpha t}) \vert \psi \rangle \Vert =0 .
\label{S1:30b}
\eeq
The two Hilbert space wave operators satisfy the Poincar\'e covariance
condition
\beq
U(\Lambda ,a) \Omega_{\alpha \pm}(H,\Phi_{\alpha},H_{\alpha}) =
\Omega_{\alpha \pm}(H,\Phi_{\alpha},H_{\alpha}) U_{\alpha} (\Lambda,a)
\qquad
U(\Lambda ,a) \Omega_{\alpha \pm}(H,\tilde{\Phi}_{\alpha},H_{\alpha}) =
\Omega_{\alpha \pm}(H,\tilde{\Phi}_{\alpha},H_{\alpha}) U_{\alpha} (\Lambda,a)
\label{S1:31}
\eeq
where, unlike (\ref{S1:28}), $(\Lambda,a)$ are not restricted to the
kinematic subgroup.

The purpose of the injection operators in Haag-Ruelle scattering is to
define the asymptotic boundary conditions.  They are constructed so
they behave like creation operators in asymptotically separated
regions.  They replace the free-particle asymptotic states of ordinary
scattering theory by states in the Hilbert spaces of the field theory
that only behave like a system of non-interacting particles in the asymptotic region.
The two injection operators defined in appendix \ref{inj} both have this
property.  They differ in which variable enforces the one-particle
mass-shell delta function when applied the vacuum. 
This means that as operators
\beq
\Omega_{\alpha \pm}(H,{\Phi}_{\alpha},H_{\alpha})=
\Omega_{\alpha \pm}(H,\tilde{\Phi}_{\alpha},H_{\alpha})
\label{S1:32}
\eeq
if they are evaluated in the same one-particle basis.   This identity 
is equivalent to the requirement
\beq
0=
\lim_{t \to \pm \infty} \Vert e^{iHt} ({\Phi}_{\alpha} - \tilde{{\Phi}}_{\alpha})
e^{-iH_{\alpha t}} \vert \psi_{\alpha} \rangle  \Vert =
\lim_{t \to \pm \infty} \Vert ({\Phi}_{\alpha} - \tilde{{\Phi}}_{\alpha})
e^{-iH_{\alpha t}} \vert \psi_{\alpha} \rangle  \Vert 
\label{S1:33}
\eeq
which is precisely the condition that the injection operators
agree asymptotically.

The scattering operator for scattering from channel
$\alpha$ to channel $\beta$  is
\beq
S_{\beta,\alpha}:{\cal H}_{\alpha} \to {\cal H}_{\beta}: =
\Omega^{\dagger}_{\beta +}(H,\Phi_\beta,H_\beta)
\Omega_{\alpha -}(H,\Phi_\alpha,H_\alpha).
\label{S1:34}
\eeq
It follows from the identity (\ref{S1:33}) that
$S_{\alpha\beta}$ is independent of the choice of injection operator.
It follows from (\ref{S1:31}) that  
\beq
U_{\beta}(\Lambda,a)S_{\beta,\alpha} =
S_{\beta,\alpha}U_{\alpha}(\Lambda,a) .
\label{S1:35}
\eeq

This can be extended to all channels, ${\cal A}$, by defining
the asymptotic Hilbert space as the direct sum of all
channel Hilbert spaces ${\cal H}_{\alpha}$ for $\alpha \in {\cal A}$,
where ${\cal A}$ is defined to include the vacuum and one-particle channels
as well as the scattering channels:
\beq
{\cal H}_{\cal A}:= \oplus_{\alpha \in {\cal A}} {\cal H}_{\alpha}. 
\label{S1:36}
\eeq
The asymptotic unitary representation of the Poincar\'e group is defined
on ${\cal H}_{\cal A}$ by
\beq
U_{\cal A}(\Lambda, a) = \oplus_{\alpha \in {\cal A}} U_{\alpha}(\Lambda, a).
\label{S1:37}
\eeq

Multi-channel injection operators are defined as the sum of all channel
injection operators
\beq
\Phi_{\cal A}:= \sum_{\alpha \in {\cal A}} \Phi_{\alpha}
\qquad \mbox{or} \qquad 
\tilde{\Phi}_{\cal A}:= \sum_{\alpha \in {\cal A}} \tilde{\Phi}_{\alpha}
\label{S1:38}
\eeq
where $\Phi_{\alpha}:{\cal H}_{\alpha} \to {\cal H}$.
These can be used to define multi-channel wave operators
\beq
\Omega_{{\cal A}  \pm} (H,\Phi_{\cal A},H_{\cal A})
=
\tilde{\Omega}_{{\cal A}  \pm} (H,\tilde{\Phi}_{\cal A},H_{\cal A})
\label{S1:39}
\eeq
by the strong limits
\beq
\lim_{t \to \pm\infty}
\Vert (\Omega_{{\cal A} \pm} (H,\Phi_{\cal A},H_{\cal A}) -
e^{iHt} \Phi_{\cal A} e^{-i H_{\cal A} t} \vert \psi \rangle \Vert =0
\label{S1:40}
\eeq
or
\beq
\lim_{t \to \pm\infty}
\Vert (\Omega_{{\cal A} \pm} (H,\tilde{\Phi}_{\cal A},H_{\cal A}) -
e^{iHt} \tilde{\Phi}_{\cal A} e^{-i H_{\cal A} t} \vert \psi \rangle \Vert =0
\label{S1:41}
\eeq
where $H_{\cal A}:= \sum_{\alpha \in {\cal A}} H_{\alpha}\Pi_{\alpha}$ and
$\Pi_{\alpha}$ is the projection on the subspace ${\cal H}_{\alpha}$
of ${\cal H}_{\cal A}$.  The
assumed asymptotic completeness of the scattering theory means that
the wave operators (including the vacuum and one-body channels) are
unitary mappings from the asymptotic Hilbert space, ${\cal H}_{\cal
  A}$, to the Hilbert, ${\cal H}$, space of the quantum field theory.

With this definition (\ref{S1:31}) becomes
\beq
U(\Lambda ,a) \Omega_{{\cal A} \pm} (H,{\Phi}_{\cal A},H_{\cal A})=
\Omega_{{\cal A}\pm} (H,{\Phi}_{\cal A},H_{\cal A}) U_{{\cal A}} (\Lambda,a)
\label{S1:42}
\eeq
which also holds with $\Phi_{\cal A}$ replaced by $\tilde{\Phi}_{\cal A}$. 

The multi-channel scattering operator is
\beq
S(H,\Phi_{\cal A},{\cal H}_{\cal A})=
\Omega_+^{\dagger}(H,\Phi_{\cal A},{\cal H}_{\cal A})
\Omega_-(H,\Phi_{\cal A},{\cal H}_{\cal A})
\label{S1:43}
\eeq
where it was constructed to be the identity on the vacuum and one
particle states.

Poincar\'e invariance of the multi-channel
scattering operator on ${\cal H}_{\cal A}$ follows from the intertwining
property of the wave operators 
(\ref{S1:42}) \cite{simon}
\[
U_{\cal A}(\Lambda,a) S(H,\Phi_{\cal A},{\cal H}_{\cal A})=
U_{\cal A}(\Lambda,a) \Omega_+^{\dagger}(H,\Phi_{\cal A},{\cal H}_{\cal A})
\Omega_-(H,\Phi_{\cal A},{\cal H}_{\cal A})=
\]
\[
\Omega_+^{\dagger}(H,\Phi_{\cal A},{\cal H}_{\cal A})
U(\Lambda ,a)\Omega_-(H,\Phi_{\cal A},{\cal H}_{\cal A})=
\]
\beq
\Omega_+^{\dagger}(H,\Phi_{\cal A},{\cal H}_{\cal A})
\Omega_-(H,\Phi_{\cal A},{\cal H}_{\cal A})
U_{\cal A}(\Lambda,a) =
S(H,\Phi_{\cal A},{\cal H}_{\cal A})
U_{\cal A}(\Lambda,a).
\label{S1:44}
\eeq
At this point all that has been demonstrated is that the same
scattering theory is obtained by using injection operators that agree
asymptotically but commute with different kinematic subgroups.  This
will be used in the next section to construct unitarily equivalent
light-front and instant-form representations of the field theory that
preserve the scattering matrix.

\section{Construction}

The next step is to use the kinematic symmetries of the injection operators
to show that the
Hamiltonian $H$ in the expressions for the wave operators
can be replaced by the light front Hamiltonian, $P^-$.  The relations
\beq
H= P^-+P^3 \qquad H = {1 \over 2}(P^++P^-)
\label{S2:1}
\eeq
will be used in what follows.
In the first case note because
$P^3{\Phi}_{\cal A}={\Phi}_{\cal A} P^3_{\cal A}$
it follows that
\beq
e^{iHt}\Phi_{{\cal A}}e^{-iH_{\cal A}} = 
e^{i(P^-+P^3)t}\Phi_{{\cal A}}e^{-i(P^-_{\cal A}+ P^3_{\cal A})t}=
e^{iP^-t}\Phi_{{\cal A}}e^{-iP^-_{\cal A}t}.
\label{S2:2}
\eeq
Similarly since $P^+\tilde{\Phi}_{\cal A}=\tilde{\Phi}_{\cal A} P^+_{\cal A}$ 
\beq
e^{iHt}\tilde{\Phi}_{{\cal A}}e^{-iH_{\cal A}} = 
e^{i(P^-+P^+)t/2}\tilde{\Phi}_{{\cal A}}e^{-i(P^-_{\cal A}+ P^+_{\cal A})t/2}=
e^{iP^-t/2}\tilde{\Phi}_{{\cal A}}e^{-iP^-_{\cal A}t/2}.
\label{S2:3}
\eeq
This means that
the wave operators constructed using $H$ are identical
to the ones using $P^-$ with both injection operators.
Combining these results with (\ref{S1:39}) gives the following
identifications
\beq
\Omega_{\pm}(H,\Phi_{\cal A},H_{\cal A}) =
\Omega_{\pm}(H,\tilde{\Phi}_{\cal A},H_{\cal A}) =
\Omega_{\pm}(P^-,\Phi_{\cal A},P^-_{\cal A}) =
\Omega_{\pm}(P^-,\tilde{\Phi}_{\cal A},P^-_{\cal A}). 
\label{S2:4}
\eeq
The final step in the construction is to introduce two
unitary operators $V_F$ and $V_I$ that act on 
the field theory Hilbert space ${\cal H}$ with the following
properties:
\beq
V_F \vert 0 \rangle = V_I \vert 0 \rangle = \vert 0 \rangle
\label{S2:5}
\eeq
\beq
[U(\Lambda_{K_I},a_{K_I}),V_I]=0 \qquad [U(\Lambda_{K_F},a_{K_F}),V_F]=0   
\label{S2:6}
\eeq
\beq
\lim_{t \to \pm \infty}
\Vert (\Phi_{\cal A} - V_I\Phi_{\cal A}) e^{-i H_{\cal A}t}\vert \psi_{\alpha} \rangle \Vert =0
\label{S2:7}
\eeq
\beq
\lim_{t \to \pm \infty}
\Vert  (\tilde{\Phi}_{\cal A} - V_F\tilde{\Phi}_{\cal A}) e^{-i H_{\cal A}t}
\vert \psi_{\alpha} \rangle \Vert =0 
\label{S2:8}
\eeq
where $(\Lambda_{K_I},a_{K_I})$ and $(\Lambda_{K_F},a_{K_F})$ are in the
instant-form and light-front kinematic subgroups respectively. 
Note that for equations (\ref{S2:7} and \ref{S2:8}) to hold the
asymptotic Hamiltonian must have a non-trivial absolutely continuous
mass spectrum.  While this is true on the scattering channel subspaces,
it is not true for the vacuum or one-particle states, where
$e^{-i H_{\cal A}t}$ either factors out because the energy depends on the
mass and kinematic variables or is $1$ in the case of the vacuum.
In these cases (\ref{S2:8}) becomes
\beq
\Vert (\tilde{\Phi}_{\cal A} - V_F\tilde{\Phi}_{\cal A}) \vert \psi_{\alpha} \rangle \Vert 
=0.
\eeq
In the construction of this section the operators $V_I$ and $V_F$ were
chosen to leave the vacuum unchanged.  While this is not necessary,
this choice gives S-matrix equivalent light-front and instant-form
field theories with the same vacuum vectors.

The next step is to define two new unitary representations of the Poincar\'e
group on the Hilbert space ${\cal H}$ of the quantum field theory by:
\beq
U_F(\Lambda ,a):=V_F U(\Lambda,a) V_F^{\dagger}
\qquad
U_I(\Lambda ,a):=V_I U(\Lambda,a) V_I^{\dagger}
\label{S2:9}
\eeq
with the corresponding dynamical generators:
\beq
H_F:=V_F H V_F^{\dagger} \qquad  H_I:=V_I H V_I^{\dagger}
\label{S2:10}
\eeq
\beq
P^-_F := V_F P^- V_F^{\dagger}  \qquad P^-_I := V_I P^- V_I^{\dagger}.  
\label{S2:11}
\eeq
It follows from (\ref{S2:6}) that
\beq
U_I(\Lambda_{K_I} ,a_{K_I}) =U(\Lambda_{K_I},a_{K_I}) 
\qquad
U_F(\Lambda_{K_F} ,a_{K_F}):= U(\Lambda_{K_F},a_{K_F})
\label{S2:12}
\eeq
where $(\Lambda_{K_I} ,a_{K_I})$ is an instant-form kinematic  Poincar\'e
transformation and $(\Lambda_{K_F} ,a_{K_F})$ is a front-form kinematic
transformation.

In Appendix \ref{seq} it is  shown \cite{Ekstein:1960} that
as a consequence of
(\ref{S2:7}) and (\ref{S2:8}) that  
\beq
\Omega_{\pm}(H_I,\Phi_{\cal A},H_{\cal A}) =
V_I \Omega_{\pm}(H,\Phi_{\cal A},H_{\cal A})
\label{S2:13}
\eeq
and
\beq
\Omega_{\pm}(H_F,\tilde{\Phi}_{\cal A},H_{\cal A}) = V_F \Omega_{\pm}(H,\tilde{\Phi}_{\cal A},H_{\cal A}).
\label{S2:14}
\eeq
  
Taken together with (\ref{S2:4}) gives:  
\beq
\Omega_{\pm}(H_I,\Phi_{\cal A},H_{\cal A}) =
V_I \Omega_{\pm}(H,\Phi_{\cal A},H_{\cal A}) =
V_I \Omega_{\pm}(H,\tilde{\Phi}_{\cal A},H_{\cal A})=
V_I \Omega_{\pm}(P^-,\tilde{\Phi}_{\cal A},P^-_{\cal A})=
V_IV_F^{\dagger} \Omega_{\pm}(P_F^-,\tilde{\Phi}_{\cal A},P^-_{\cal A}).
\label{S2:15}
\eeq
Since (\ref{S2:15}) holds for the same $V_IV_F^{\dagger}$ for both time limits
it follows that 
\[
S(H_I,\Phi{\cal A},H_{\alpha}) =
\Omega^{\dagger}_{+}(H_I,\Phi_{\cal A},H_{\cal A})
\Omega_{-}(H_I,\Phi_{\cal A},H_{\cal A}) =
\]
\[
\Omega^{\dagger}_{+}(H_F,\tilde{\Phi}_{\cal A},H_{\cal A})
V_FV_I^{\dagger} V_IV_F^{\dagger}
\Omega_{-}(H_F,\tilde{\Phi}_{\cal A},H_{\cal A}) =
\]
\[
\Omega^{\dagger}_{+}(P^-_F,\tilde{\Phi}_{\cal A},P^-_{\cal A})
\Omega_{-}(P^-_F,\tilde{\Phi}_{\cal A},P^-_{\cal A}) =
\]
\beq
S(P^-_F,\tilde{\Phi}_{\cal A},P^-_{\cal A}).
\label{S2:17}
\eeq
This means both representations of the dynamics result in the same scattering
operators on the asymptotic Hilbert space.
Next consider the two unitary representations of
the Poincar\'e group, $U_I(\Lambda,a)$ and $U_F(\Lambda,a)$
defined above.
For the first one the operators $\{\mathbf{P},s,s_z\}$
are mutually commuting self-adjoint functions of the instant-form
kinematic generators of the representation $U_I(\Lambda,a)$.  For the
second one the operators $\{\tilde{\mathbf{P}},s_{z_f}\}$
are mutually commuting self-adjoint functions of the
front-form kinematic generators of the representation
$U_F(\Lambda,a)$.

It is possible to construct a basis for $U_I(\Lambda,a)$
consisting of the eigenvalues of $\{\mathbf{P},s,s_{zI}\}$ and
some additional kinematically invariant commuting observables $x_I$.
Similarly it is possible to construct a basis for
$U_F(\Lambda,a)$ consisting of the eigenvalues of $\{\tilde{\mathbf{P}},s_{zF}\}$ and some additional kinematically invariant commuting observables $x_F$.

It follows that in these bases wave functions have the form
\beq
\langle \mathbf{p}, s,\mu_I ,x_I \vert \psi \rangle 
\qquad \mbox{or} \qquad
\langle \tilde{\mathbf{p}} ,\mu_F ,x_F \vert \psi \rangle .
\label{S2:19}
\eeq
Matrix elements of the dynamical operators
are non-trivial matrices in the $``x''$ variables: 
\beq
\langle \mathbf{p}, s,\mu_I ,x_I \vert H_I \vert
\mathbf{p}', s',\mu_I' ,x_I'  \rangle =
\delta (\mathbf{p}-\mathbf{p}') \delta_{ss'}\delta_{\mu_I \mu_I'} 
\langle x_I \Vert H_I(\mathbf{P},s) \Vert x_I' \rangle
\label{S2:20}
\eeq
and
\beq
\langle \tilde{\mathbf{p}},\mu_F ,x_F \vert P^-_F \vert
\tilde{\mathbf{p}}',\mu_F' ,x_F'  \rangle =
\delta (\tilde{\mathbf{p}}-\tilde{\mathbf{p}}') \delta_{\mu_F \mu_F'} 
\langle x_F \Vert P^-_F(\tilde{\mathbf{P}},\mu_F) \Vert x_F' \rangle .
\label{S2:21}
\eeq

The eigenvalue problems for the dynamical operators have the forms
\beq
\sum\int \langle x_I \Vert H_I(\mathbf{P},s) \Vert x_I' \rangle
dx_I' \langle \mathbf{p}, s,\mu_I ,x_I \vert \psi \rangle =
E(\mathbf{P},s) \langle \mathbf{p}, s,\mu_I ,x_I \vert \psi \rangle
\label{S2:22}
\eeq
and
\beq
\sum\int 
\langle x_F \Vert P^-_F(\tilde{\mathbf{P}},\mu_F) \Vert x_F' \rangle
dx_F' \langle \tilde{\mathbf{p}},\mu ,x_F' \vert \psi \rangle =
P^-(\tilde{\mathbf{P}},\mu_F) \langle \tilde{\mathbf{p}},\mu_F ,x_F
\vert \psi \rangle .
\label{S2:23}
\eeq
The matrices
$\langle x_I \Vert H_I(\mathbf{P},s) \Vert x_I' \rangle$ or
$\langle x_F \Vert P_F^-(\tilde{\mathbf{P}},\mu) \Vert x_F' \rangle$
must be diagonalized in order to compute
dynamical Poincar\'e transformations.  On the other hand
kinematic transformations can be computed by applying the inverse
kinematic transformation to basis vectors:
\[
\langle \mathbf{p}, s,\mu ,x_I \vert U_I(R,\mathbf{a})\vert  \psi \rangle =
\langle \psi \vert  U_I^{\dagger}(R,\mathbf{a}) \vert
\mathbf{p}, s,\mu ,x_I  \rangle^* =
\]
\beq
e^{i \mathbf{a} \cdot \mathbf{p}} \langle 
R^{-1}\mathbf{p}, s,\nu ,x_I \vert \psi  \rangle
D^s_{\mu \nu}(R).
\label{S2:24}
\eeq
Similarly in the light-front case for light-front preserving
boosts and translations:
\[
\langle \tilde{\mathbf{p}},\mu ,x_F \vert U_F(\Lambda_{KF},a_{KF})\vert  \psi \rangle =
\langle \psi \vert  U_F^{\dagger}(\Lambda_{KF},a_{KF}) \vert
\tilde{\mathbf{p}},\mu ,x_F \rangle^* = 
\]
\beq
e^{i \tilde{\mathbf{a}} \cdot \tilde{\mathbf{p}}}
\langle \tilde{\pmb{\Lambda}}^{-1}_k p,\mu ,x_F \vert  \psi \rangle
\sqrt{{ (\Lambda^{-1}_k p)^+ \over p^+}}
\label{S2:25}
\eeq
and for rotations about the $z$ axis
\[
\langle \tilde{\mathbf{p}},\mu ,x_F \vert U_F(R_z(\phi),0) \psi \rangle =
\langle \psi \vert  U_F^{\dagger}(R_z(\phi)) \vert
\tilde{\mathbf{p}},\mu ,x_F \rangle^* = 
\]
\beq
\langle \tilde{\mathbf{R}}_z(\phi)^{-1}p,\mu ,x_F \vert  \psi \rangle
s^{i \mu \phi}.
\label{S2:26}
\eeq
Equation (\ref{S2:24}) shows that $U_I(\Lambda,a)$ has an instant-form kinematic subgroup
while equations (\ref{S2:25}) and (\ref{S2:26}) show that
$U_F(\Lambda,a)$
has a light-front
kinematic subgroup.  In addition the two representations are related
by a unitary transformation:  
\beq
U_F(\Lambda,a)= V_FV_I^{\dagger}U_I(\Lambda,a)V_IV_F^{\dagger}
\label{S2:27}
\eeq
\beq
\vert 0 \rangle_F =  V_FV_I^{\dagger} \vert 0 \rangle_I = \vert 0 \rangle_I .
\label{S2:28}
\eeq
In this construction it was chosen
so that it preserves the vacuum.

If follows from (\ref{S2:17}) that both representations give the same unitary scattering
operators on the asymptotic Hilbert space ${\cal H}_{\cal A}$.  In addition
because $V_I$ and $V_f$ are kinematically invariant (\ref{S2:6}),
the injection operators satisfy
\beq
U_I(\Lambda_{KI} ,a_{KI}) \Phi_{\cal A} =
\Phi_{\cal A}U_{\cal A}(\Lambda_{KI} ,a_{KI})
\label{S2:29}
\eeq
for the instant-form kinematic subgroup and 
\beq
U_F(\Lambda_{KF} ,a_{KF}) \tilde{\Phi}_{\cal A} =
\tilde{\Phi}_{\cal A}U_{\cal A}(\Lambda_{KF} ,a_{KF})
\label{S2:30}
\eeq
for the light-front kinematic subgroup.

The results can be summarized by noting that it is possible to
choose bases in the field theory Hilbert space that transform
covariantly under either kinematic subgroup.  In both cases a
free dynamics is not assumed. The two representations are related
by $S$-matrix preserving unitary transformations.  Both representation have
the same vacuum and one-particle subspaces.  There are no bare particles
in these representations.

This construction did not assume the existence of a free dynamics
on the Hilbert space of the field theory. 
If the range of $\tilde{\Phi}$ is all of
${\cal H}$ and
\beq
W:=  (\tilde{\Phi}_{\cal A} \tilde{\Phi}_{\cal A}^{\dagger})^{-1/2} \tilde{\Phi}_{\cal A} 
\label{S2:31}
\eeq
satisfies
\beq
\lim_{t \to \pm \infty}
\Vert (W -\tilde{\Phi}_{\cal A}) e^{-i H_{\cal A}t}\vert \psi_{\alpha} \rangle \Vert =0
\label{S2:32}
\eeq
then $W$ 
is a kinematically invariant unitary mapping from
${\cal H}_{\cal A}$ to ${\cal H}$ and
$U_0(\Lambda,a) = W U_{\cal A}(\Lambda,a) W^{\dagger}$
defines a consistent ``free dynamics'' on ${\cal H}$
that satisfies
\beq
U_I(\Lambda_{KI},a_{KI}) = U_0(\Lambda_{KI},a_{KI})
\label{S2:33}
\eeq
with a similar relation in the light-front case. If these
representation exist, they are not unique since they depend on the
choice of injection operator.

\section{Spontaneous Symmetry Breaking}

One of the puzzling aspects of light-front quantum field theory
is how spontaneous symmetry breaking is realized.  The
signal for spontaneous symmetry breaking is a conserved local current,
$j^{\mu}(x)$, that arises from the symmetry and the presence of a 0 mass
Goldstone boson in the spectrum.  For a local scalar field theory 
the following condition \cite{coleman}
\beq
\lim_{R \to \infty} \langle 0 \vert [Q_R, \phi(y)] \vert 0 \rangle \not=0
\label{SB:1}
\eeq
implies the existence of a 0 mass particle.  Here 
\beq
\langle 0 \vert [Q_R, \phi(y)] \vert 0 \rangle :=
\langle 0 \vert [ \int d\mathbf{x} \chi_R(\vert \mathbf{x} \vert)
j^0(\mathbf{x},t), \phi(y)] \vert 0 \rangle
\label{SB:2}
\eeq
where $\chi_R(\vert \mathbf{x} \vert)$ is a smooth function that
that is 1 for $\vert \mathbf{x}\vert <R$ and $0$ for $\vert \mathbf{x}\vert
>R+\epsilon$ for some finite positive $\epsilon$.  The cutoff
function $\chi_R$ ensures that the integral converges for
large $\vert\mathbf{x}\vert$.  The commutator vanishes for
$x-y$ space-like.  For fixed $y$ and $t$ and sufficiently large
$R$, if $\vert \mathbf{x}\vert >R$ then $x-y$ is space-like and
the commutator vanishes.  In this case
\beq
\langle 0 \vert [Q_R, \phi(y)] \vert 0 \rangle :=
\langle 0 \vert [ \int d\mathbf{x} \chi_R(\vert \mathbf{x} \vert)
j^0(\mathbf{x},t), \phi(y)] \vert 0 \rangle =
\int d\mathbf{x} \langle 0 \vert [ 
j^0(\mathbf{x},t), \phi(y)] \vert 0 \rangle .
\label{SB:3}
\eeq
{\it This does not require that the charge
operator to exists. } 

Current conservation implies
\beq
\langle 0 \vert [ \int d\mathbf{x}
\partial_{\mu} j^\mu(\mathbf{x},t)
, \phi(y)] \vert 0 \rangle = 0 .
\label{SB:4}
\eeq
Inserting a complete set of intermediate states gives
\[
0=\sum \int d\mathbf{x} \left(
\langle 0 \vert \partial_{\mu} j^\mu(\mathbf{x},t)
\vert p ,n \rangle 
{d\mathbf{p} \over 2p_n^0}  \langle p,n \vert \phi(y) \vert 0 \rangle
- \langle 0 \vert  \phi(y)
\vert p ,n \rangle {d\mathbf{p} \over 2p_n^0}
\langle p,n \vert \partial_{\mu}
j^\mu(\mathbf{x},t)  \vert 0 \rangle\right )=
\]
\beq
\sum\int d\mathbf{x} \left(
\langle 0 \vert \partial_{\mu} j^\mu(\mathbf{0 },0)
\vert p_r ,n \rangle {d\mathbf{p} \over 2p_n^0} \langle p_r,n \vert \phi(0) \vert 0 \rangle
e^{i p\cdot (x-y)}
- \langle 0 \vert  \phi(0)
\vert p_r ,n \rangle {d\mathbf{p} \over 2p_n^0} \langle p_r,n \vert \partial_{\mu}
j^\mu(\mathbf{0},0)  \vert 0 \rangle
e^{i p\cdot (y-x)}
\right )
\label{SB:5}
\eeq
where Poincar\'e covariance has been used to remove the non-trivial
$x,y$ and $p$ dependence from the matrix elements. $p_r$ is
the constant rest four momentum for massive states and a constant light-like
vector for massless states.  For a scalar field theory the vacuum
expectation value of the current vanishes so the vacuum does
not appear as an intermediate state.  The Lehmann weights that appear in
this matrix element
\beq
\sigma (m_n)m_n^2 = 
\langle 0 \vert \partial_{\mu} j^\mu(\mathbf{0},0)
\vert p_r ,n \rangle \langle p_{nr},n \vert \phi(0) \vert 0 \rangle
\label{SB:6}
\eeq
and
\beq
\sigma^*(m_n)m_n^2 =
\langle 0 \vert  \phi(0)
\vert p_{nr} ,n \rangle \langle p_{nr},n \vert \partial_{\mu}
j^\mu(\mathbf{0},0)  \vert 0 \rangle
\label{SB:7}
\eeq
are functions of the invariant mass of the intermediate states.
The factor $m_n^2=-p_n^2$ is due to the $\partial_{\mu} j^\mu(\mathbf{0},0)$
assuming that the current is a local function of the scalar field.

Current conservation requires that this quantity vanishes which
follows if either $\sigma(m_n)=0$ or $m_n^2=0$.
Inserting intermediate states in (\ref{SB:2}) gives a similar result
with $\sigma (m_n)m_n^2$ replaced by  
$p_{nr}^0\sigma (m_n)$ where $p_{nr}^0$ is a non-zero constant.

Since this does not include the $m^2$ factor,
it vanishes by current conservation unless $\sigma(m_n)$
contains a $\delta (m_n)$.  Thus for (\ref{SB:1}) to be non-vanishing 
the sum over intermediate states must have a contribution from
a 0 mass particle.   This is the non-perturbative
form of Goldstone's theorem.
%
%

In \cite{coleman} Coleman avoids directly
computing the charge operator.  The current is an operator valued
distribution, so there is no reason to expect that integrating against
the constant $1$ will result in a well defined operator.  There is
another reason for this.  If the integrals that define the charge
operator converge and there is no contribution to the current on the
surface at infinity, then the charge operator will be translationally
invariant and time independent.  This means that it would commute with
the four momentum operator.  It follows that even if $Q_I \vert 0
\rangle$ is not $0$, it is eigenstate of the four momentum, with 0
eigenvalue.  While the Goldstone boson has 0 mass, its momentum cannot
be 0, which means that a translationally invariant time-independent
charge operator cannot couple the vacuum to the Goldstone boson.

In the light-front representation it is still possible to define
$Q_{IR}(t)$ and the analysis above still applies.  The assumption that
the current is a local function of the fields means that the
light-front dynamics must be used to get its value on the fixed-time
surface where the locality of the theory can be utilized to control the
volume integral.

While the non-vanishing of (\ref{SB:1}) is a sign of spontaneous
symmetry breaking, the argument used above does not work when it is
applied to the light-front charge.  The problem is that there is no
compact region on the light front where outside of that region $(x-y)$
is always space-like for fixed $y$ and $x^+=0$.  This is true even if
the light-front charge is computed in the instant-form representation.
If the integrals that define the light-front charge exist and there
are no contributions from the divergence theorem, the situation is
similar to the instant-form case - the charge operator will commute
with the four momentum which means that the light-front charge applied
to the vacuum is an eigenstate of the four momentum with 0 eigenvalue,
which by the argument used above means that the charge operator cannot
couple the vacuum to the Goldstone boson.

\section{Miscellaneous remarks}

This section provides a brief discussion comparing
the structure of the light-front and instant-form representations
discussed in this work to the corresponding relation in relativistic
quantum mechanics of a finite number of degrees of freedom.

The starting point of this work is the assumed existence of a local
field theory with the expected properties.  Normally in an instant or
front-form representation of a quantum field theory ``solving the
field theory'' means constructing a unitary transformation that
decomposes the Hilbert space into a direct integral of irreducible
representations.  This is the relativistic analog of diagonalizing the
Hamiltonian.  This is also true in relativistic quantum theories of a
finite number of degrees of freedom, except in that case the
interacting and non-interacting theories are formulated on the same
representation of the Hilbert space.  In the finite number of degree
of freedom case different choices of independent variables in each
irreducible subspace are related by a unitary change of variables.
This defines a unitary, S-matrix preserving transformation relating
the two forms of
dynamics.  It is non-trivial because there is a different (mass-dependent)
variable change on each irreducible subspace.
The light-front and instant-form bases have the property
that matrix elements of the unitary representation of the kinematic
subgroups in each case are mass independent.

In this work this process is reversed; the direct integral is assumed
to be a property of a quantum field theory with all of the expected
properties.  The general class of S-matrix preserving unitary
transformations is used to construct equivalent representations of the
dynamics.  The freedom to choose different $S-$matrix preserving
unitary transformations was exploited to choose S-matrix preserving
unitary transformations that commute with the instant form or
light-front kinematic subgroups of the theory, which resulted in
equivalent light-front or instant-form representations of the field
theory.

The unitarity of these transformations means that they must preserve
the spectrum of the mass operator.  This means that point-spectrum
eigenvalues of the mass operator must be the same in all three
representations.  The unitary transformation must also leave the
vacuum energy or the mass gap unchanged.  Preserving the $S$-matrix
means that acceptable unitary transformations must be short-range
perturbations of the identity in the sense that they do not require a
modification of scattering asymptotic conditions.   The unitary
transformations that satisfy all of these conditions are not unique
and are discussed in appendix (\ref{seq}).


The construction of equivalent instant-form and light-front unitary
representations of the Poincar\'e group in this paper is essentially
identical to the corresponding construction in the finite number of
degree of freedom case
\cite{Sokolov:1977im}\cite{Keister:1996bd}\cite{Polyzou:19892b}\cite{Polyzou:2002cp}.
In that case, when there is physical particle production,
the particles all have their
physical masses and there are no vertex interactions that can change
masses \cite{Polyzou:2003dt}. In that case the simplest interactions that change
particle number are $2 \leftrightarrow 3$ interactions.

Given the spectrum of one-particle states, the mass spectrum of the
field theory Hamiltonian includes the one-particle masses and
continuous eigenvalues starting at the two-particle thresholds.  This is
identical to the mass spectrum of a sum of free-field Hamiltonians with
physical particle masses.  The conclusion of Haag's theorem
\cite{Wightman:1980} is that a spectrally-equivalent free field
theories are defined on inequivalent representation of the Hilbert
space.  This why there are no free multiparticle states in the field
theory and is the motivation for using the two-Hilbert space
representation in order to formulate the scattering asymptotic
conditions.

In the absence of free multi-particle states in the theory, the notion
of many-particle interactions needs to be defined.  One possibility is
to define interactions using matrix elements in states constructed by
applying products of the operators that create one-particle states out
of vacuum to the vacuum.  Matrix elements of the Hamiltonian in these
states have two properties, the first is that they become products of
one-particle states as the particles are asymptotically separated.  In
this representation, because the one-particle states are point spectrum
eigenstates
of the Hamiltonian, the Hamiltonian cannot couple them to the
continuum.  This means that in this representation a ``production
vertex'' cannot couple to a multiparticle scattering state.  This is
one way of saying that stable particle cannot decay, however particle
production is possible since interactions that couple ``two-particle
states'' to ``three-particle states'', etc. are possible in this
representation.

A final comment concerns vacuum polarization in this representation.
The physical vacuum is orthogonal to all bound and scattering eigenstates
of the Hamiltonian, but operators that couple the vacuum to other states
in the Hilbert space are possible and are needed to describe local vacuum
excitations.

\section{Analysis and Conclusions}

This work provides one way to understand Dirac's forms of dynamics in a
non-perturbative setting.  This approach assumed the
existence of a local quantum field theory with all of the expected
properties.  Using this assumption unitarily equivalent representations with the
same scattering operators were constructed.  Bases that transform
covariantly with respect to light-front and instant-form kinematic
subgroups where constructed for each of these equivalent
representations of the dynamics.  Kinematic Poincar\'e transformations
on these wave functions could be computed without diagonalizing the
dynamical operators.  One aspect of the construction used
in this work is that both representations of the
dynamics leave the vacuum and one-particle states
unchanged and both representations give the same
Poincar\'e invariant scattering operator.  While
preserving the vacuum states is not required, 
the construction demonstrates that it is a consistent assumption.

A key difference with the perturbative approach is the non-existence
of an underlying free-field dynamics on the Hilbert space of the field
theory.  If the injection operators have full range on the Hilbert
space of the field theory and satisfy (\ref{S2:32}) there is the
analog of a free dynamics on the Hilbert of the field theory.  If such
a representation exists, it cannot be a local free field theory with
physical masses.

While the representations discussed in this analysis are not useful
for applications, the results support the conclusions of most of the
quoted references that the two formulations of the theory are
equivalent.  The vacuum is the same up to a unitary transformation and
can be taken as the same in both representations.  While the
light-front vacuum is kinematically invariant, it is also rotationally
invariant, which is an additional dynamical constraint on the
light-front vacuum that separates it from the free-field Fock
vacuum.  By starting with the assumed solution to the field theory there
are no issues with renormalization, rotational covariance or 0 modes.
These issues need to be addressed in applications.
If there is spontaneous
symmetry breaking, the non-perturbative condition for the presence of
a Goldstone boson in the mass spectrum is the same in both
representations, but it does not couple to the vacuum with a space-time
translationally invariant charge operator, whether it is on the
fixed-time surface or the light-front hyperplane.

The author would like to thank the referee for some useful suggestions.

\appendix
\section{Scattering equivalences}
\label{seq}
The two-Hilbert space multi-channel wave operators are
defined by the strong limits where the $\Phi_{\cal A}$ are
two-Hilbert space injection operators that map the
asymptotic Hilbert space ${\cal H}_{\cal A}$ to the Hilbert space
${\cal H}$ of the quantum field theory:
\beq
\Omega_\pm(H,\Phi_{\cal A},H_{\cal A}) =
s-\lim_{t \to \pm \infty}
e^{iHt} \Phi_{\cal A} e^{-iH_{\cal A}t}. 
\label{A1:1}
\eeq
The wave operators are assumed to exist and be Poincar\'e
invariant in the sense
\beq
U(\Lambda ,a)  \Omega_\pm^{\dagger}(H,\Phi_{\cal A},H_{\cal A})=
\Omega_\pm^{\dagger}(H,\Phi_{\cal A},H_{\cal A})U_{\cal A} (\Lambda ,a).
\label{A1:2}
\eeq
where $U_{\cal A} (\Lambda ,a)$ is the natural representation of the
Poincar\'e group on ${\cal H}_{\cal A}$ that has the form of a direct
sum of tensor products of irreducible representations.  The set of
channels ${\cal A}$ is assumed to include the vacuum channel, and
one-particle channels in addition to multi-particle scattering
channels.  The wave operators are assumed to be asymptotically complete
unitary mappings from asymptotic Hilbert space ${\cal H}_{\cal A}$ to
the Hilbert space ${\cal H}$ of the field theory.  

The scattering operator is defined by 
\beq
S(H,\Phi,H_{\cal A}) =
\Omega_+^{\dagger}(H,\Phi_{\cal A},H_{\cal A})
\Omega_-(H,\Phi_{\cal A},H_{\cal A}).
\label{A1:3}
\eeq
Consider two identical scattering operators based on different
Hamiltonians
\beq
S(H,\Phi_{\cal A},H_{\cal A}) =S(H',\Phi_{\cal A}',H_{\cal A}) 
\label{A1:4}
\eeq
It follows from the definitions and unitarity of the
wave operators that
\beq
W:= \Omega_+(H',\Phi'_{\cal A},H_{\cal A})
\Omega_+^{\dagger}(H,\Phi_{\cal A},H_{\cal A})=
\Omega_-(H',\Phi'_{\cal A},H_{\cal A})
\Omega_-^{\dagger}(H,\Phi_{\cal A},H_{\cal A})
\label{A1:5}
\eeq
is a unitary operator on ${\cal H}$.
The identification of the scattering operators (\ref{A1:4}) means
that $W$ is the same for the incoming ($t \to +\infty$) or outgoing
($t \to -\infty$) multi-channel wave operator.
It follows that
\beq
W \Omega_\pm(H,\Phi_{\cal A},H_{\cal A})
=
\Omega_\pm(H',\Phi_{\cal A}',H_{\cal A}).
\label{A1:6}
\eeq
In addition, the intertwining property of the wave operators \cite{simon},
\[
WH'=
\Omega_\pm(H,\Phi_{\cal A},H_{\cal A})
\Omega_\pm^{\dagger}(H',\Phi_{\cal A}',H_{\cal A})H'
=
\Omega_\pm(H,\Phi_{\cal A},H_{\cal A})
H_{\cal A} \Omega_\pm^{\dagger}(H',\Phi'_{\cal A},H_{\cal A})
\]
\beq
=
H\Omega_\pm(H,\Phi_{\cal A},H_{\cal A})
\Omega_\pm^{\dagger}(H',\Phi'_{\cal A},H_{\cal A})=
HW
\label{A1:7}
\eeq
means that the two Hamiltonians are related by the unitary
transformation $W$.

It follows that
\[
\Omega_-(H',\Phi_{\cal A}',H_{\cal A}) = 
W \Omega_-(H,\Phi_{\cal A},H_{\cal A}) =
\]
\beq
\Omega_-(WHW^{\dagger} ,W\Phi_{\cal A},H_{\cal A})=
\Omega_-(H',W\Phi_{\cal A},H_{\cal A}).
\label{A1:8}
\eeq
Taking the difference of the right and left side of equation 
(\ref{A1:8})
gives
\beq
0=\lim_{t \to \pm \infty}
\Vert e^{iH't} (\Phi_{\cal A}' - W\Phi_{\cal A}) e^{-iH_{\cal A}t}\vert \psi \rangle \Vert .
\label{A1:9}
\eeq
The unitarity of $e^{iH't}$ gives 
\beq
0=\lim_{t \to \pm \infty} \Vert (\Phi_{\cal A}' - W\Phi_{\cal A})
e^{-iH_{\cal A}t}\vert \psi \rangle \Vert .
\label{A1:10}
\eeq
That this condition holds for both time limits is important.
For the vacuum and one-particle channels this condition requires
that
\beq
\Phi_{\cal A}' - W\Phi_{\cal A} =0.
\eeq

Next consider the converse.  Assume that $W$ is a unitary operator
satisfying $H'=WHW^{\dagger}$, and the scattering operators
\beq
S(H,\Phi,H_{\cal A}) =
\Omega_+^{\dagger}(H,\Phi_{\cal A},H_{\cal A})
\Omega_-(H,\Phi_{\cal A},H_{\cal A})
\label{A1:11}
\eeq
\beq
S(H',\Phi'_{\cal A},H_{\cal A}) =
\Omega_+^{\dagger}(H',\Phi'_{\cal A},H_{\cal A})
\Omega_-(H',\Phi'_{\cal A},H_{\cal A})
\label{A1:12}
\eeq
both exist.
If $W$ satisfies (\ref{A1:10}) for both time limits then the $S$
matrices are identical
\beq
S(H',\Phi'_{\cal A},H_{\cal A}) = S(H,\Phi_{\cal A},H_{\cal A}).
\label{A1:13}
\eeq
The proof follows from
\[
\Omega_\pm(H',\Phi'_{\cal A},H_{\cal A}) =
\Omega_\pm(WHW^{\dagger} ,\Phi'_{\cal A},H_{\cal A})=
W\Omega_\pm(H,W^{\dagger}\Phi'_{\cal A},H_{\cal A})=
\]
\beq
W\Omega_\pm(H,W^{\dagger}(\underbrace{\Phi'_{\cal A}- W\Phi{\cal A}}_{\to 0} + W\Phi_{\cal A}),H_{\cal A})=
W\Omega_\pm(H,\Phi_{\cal A}),H_{\cal A}).
\label{A1:14}
\eeq
It then follows that
\[
S(H',\Phi'_{\cal A},H_{\cal A})=
\Omega_+^{\dagger}(H',\Phi'_{\cal A},H_{\cal A})
\Omega_-(H',\Phi'_{\cal A},H_{\cal A})=
\]
\beq
\Omega_+^{\dagger}(H,\Phi_{\cal A},H_{\cal A})W^{\dagger}
W
\Omega_-(H,\Phi_{\cal A},H_{\cal A})=
S(H,\Phi_{\cal A},H_{\cal A}).
\label{A1:15}
\eeq
Note that unitary equivalence is not a sufficient condition for
$S$-matrix equivalence.  This is not hard to understand in the case of ordinary
quantum mechanics, where two Hamiltonians with short-range
repulsive interactions have
the same spectrum (so they are unitarily equivalent) but generally have
different $S$-matrix elements or phase shifts.

The conclusion of this section is that unitary operators $W$ that
satisfy the asymptotic condition (\ref{A1:10}) can be used to relate
Hamiltonians that have the same scattering matrix.  This asymptotic
condition means that $W$ does not
disturb the asymptotic structure of the states that define the
asymptotic condition.

The above condition applies to both quantum mechanics and quantum
field theory assuming that they have asymptotically complete scattering
operators.
The field theory generalization of
multi-particle scattering is the Haag-Ruelle formulation of scattering
which involves the strong limits used in this appendix.

\section{Two Hilbert space injection operators}
\label{inj}
This appendix discusses the construction of injection operators from operators
in the field algebra.  The starting point is to let $B$ be a function of the
fields that creates a one-body state out of the vacuum.  This means
that the completeness sum 
\beq
\langle 0 \vert B^{\dagger}B \vert 0 \rangle =
\sum_n \langle 0 \vert B^{\dagger}\vert n \rangle
\langle n\vert  B \vert 0 \rangle 
\label{A2:1}
\eeq
has one-body intermediate states (here one-body means states with
discrete positive-mass eigenvalues, so  there is no distinction between
elementary and composite one-body states).  For simplicity it is
assumed that the one-body spectrum is non-degenerate, which means
that each
one-particle state has a different mass, and each one-body mass
eigenstate has a given spin.

To isolate the operators that create the one-body states use
space-time translations to define the operator valued distributions
\beq
B(x) := e^{-i P\cdot x} B e^{-i P\cdot x} = U^{\dagger}(I,x) B U(I,x)
\label{A2:2}
\eeq
where $P^{\mu}$ is the four momentum operator.  It follows from
(\ref{A2:2}) that
\beq
{\partial B(x)\over \partial x_{\mu}}= -i [P^{\mu},B(x)].
\label{A2:3}
\eeq
The next step is to compute the Fourier transform of the operator density
$B(x)$: 
\beq
\hat{B}(q) := \int {d^4x \over (2 \pi)^2}
e^{iq\cdot x} B(x) .
\label{A2:4}
\eeq
Multiplying both sides of (\ref{A2:3}) by ${e^{iq\cdot x}\over (2 \pi)^2}$
and integrating over $d^4x$ by parts, assuming that ${B}(x)$ is an operator
valued distribution, gives
\beq
-i [P^{\mu},\hat{B}(q)]= -i q^{\mu} \hat{B}(q)
\label{A2:5}
\eeq
or 
\beq
[P^{\mu},\hat{B}(q)] = q^{\mu} \hat{B}(q).
\label{A2:6}
\eeq
It follows from (\ref{A2:6}) that $\hat{B}(q)\vert 0 \rangle$ is
either $0$ or an
eigenstate of the four momentum with eigenvalue $q^{\mu}$:
\beq
P^{\mu}\hat{B}(q) \vert 0 \rangle =
\hat{B}(q) P^{\mu}  \vert 0 \rangle =
q^{\mu} \hat{B}(q)\vert 0 \rangle=
q^{\mu} \hat{B}(q)\vert 0 \rangle .
\label{A2:7}
\eeq

Let $m$ be a discrete mass eigenvalue of $M=\sqrt{-P^2}$ and let $s$
denote the spin of the particle of mass $m$.  Next let $h(q)$ be a
smooth Lorentz invariant function of compact support in $q^2$ that is 1
when $q^2=-m^2$ and $q^0>0$, and
identically $0$
on the rest of the spectrum of intermediate states in (\ref{A2:1}) and
define
\beq
{A}^{\dagger}(q):= \hat{B}(q)h(q).
\label{A2:8}
\eeq
Additional integrations are used to define the operators:
\beq
{A}^{\dagger}(\mathbf{q}):= \int {A}^{\dagger}(q)  dq^0
\qquad
\mbox{and}
\qquad
\tilde{A}^{\dagger}(\tilde{\mathbf{q}}):= \int {A}^{\dagger}(q) dq^-/2 .
\label{A2:9}
\eeq
Because $h(q)$ is localized, these operators are distributions in the
three momentum or light-front components of the four momentum.  When
these operators are applied to the vacuum they create one-particle states of
mass $m$.  Since by assumption, there is a unique spin $s$ associated
with each discrete mass eigenvalue, the particle created out of the
vacuum also has spin $s$.

Because of the integrals over $q^0$ or $q^-$ 
it follows that (\ref{A2:6}) must be replaced by 
\beq
[\mathbf{P},{A}^{\dagger}(\mathbf{q})] = \mathbf{q} {A}^{\dagger}(\mathbf{q}) .
\qquad
[\tilde{\mathbf{P}},\tilde{A}^{\dagger}(\tilde{\mathbf{q}})] =
\tilde{\mathbf{q}} \tilde{A}^{\dagger}(\tilde{\mathbf{q}}) 
\label{A2:10}
\eeq
which when applied to the vacuum becomes
\beq
\mathbf{P}A^{\dagger}(\mathbf{q}) \vert 0 \rangle  = \mathbf{q}{A}^{\dagger} (\mathbf{q})\vert 0 \rangle
\qquad
\tilde{\mathbf{P}}\tilde{A}^{\dagger}(\tilde{\mathbf{q}}) \vert 0 \rangle  =
\tilde{\mathbf{q}}\tilde{A}^{\dagger} (\tilde{\mathbf{q}})\vert 0 \rangle .
\label{A2:11}
\eeq
This means that ${A}^{\dagger}(\mathbf{q})$ creates a one-particle
state with momentum $\mathbf{q}$, mass $m$ and spin $s$ out of the vacuum.
Similarly
\beq
\tilde{\mathbf{P}}
\tilde{A}^{\dagger}( \tilde{\mathbf{q}}) \vert 0 \rangle  =
\tilde{\mathbf{q}}{A}^{\dagger} (\tilde{\mathbf{q}})\vert 0 \rangle .
\label{A2:12}
\eeq
creates a one-particle state with light-front momentum $\tilde{\mathbf{q}}$
and mass $m$ and spin $s$ out of the vacuum.

The next step is to construct operators that create simultaneous
eigenstates of mass and spin and magnetic quantum numbers.

Define the operator 
\beq
{A}^{\dagger}_{m,s}( \mathbf{p},\mu) := \int_{SU(2)}
U(R,0){A}^{\dagger}(R\mathbf{p} )U^{\dagger}(R,0)
D^{s*}_{\mu s}(R)
dR 
\label{A2:13}
\eeq
where the integral is over the $SU(2)$ Haar measure and
$D^{s*}_{\mu s}(R)$ is the spin-$s$ $SU(2)$ Wigner function
\beq
D^{s}_{\mu s}(R) = \langle s, \mu \vert U(R)\vert s,s \rangle .
\label{A2:14}
\eeq

It follows from (\ref{A2:13}) that
\[
U(R',0){A}^{\dagger}_{m,s}(\mathbf{p},\mu)U^{\dagger}(R',0) =
\]
\beq
\int_{SU(2)}
U(R'R,0){A}^{\dagger}(R^{-1}\mathbf{p} )U^{\dagger}(R'R,0) D^{s*}_{\mu s}(R) dR .
\label{A2:15}
\eeq
Changing variables,  
$R''=R'R$ using the invariance of the Haar measure $dR = dR''$
for fixed $R'$ gives
\[
(\ref{A2:15})=\int_{SU(2)}
U(R'',0){A}^{\dagger}(R^{\prime \prime-1}R'\mathbf{p} )U^{\dagger}(R'',0) dR''
D^{s*}_{\mu s}(R^{\prime -1}R^{\prime \prime})=
\]
\[
\int_{SU(2)}\sum_{\nu=-s}^s
U(R'',0)A^{\dagger}(R^{\prime \prime-1}R'\mathbf{p} )U^{\dagger}(R'',0) dR''
D^{s*}_{\mu\nu} (R^{\prime -1})D^{s*}_{\nu s}(R^{\prime \prime})=
\]
\[
\int_{SU(2)}\sum_{\nu=-s}^s
U(R'',0){A}^{\dagger}(R^{\prime \prime-1}R'\mathbf{p} )U^{\dagger}(R'',0) dR''
D^{s*}_{\nu s}(R^{\prime -1}R^{\prime \prime})
D^{s}_{\nu\mu} (R^{\prime}) =
\]
\beq
\sum_{\nu=-s}^s
{A}^{\dagger}_{(m,j)}(R' \mathbf{p},\nu) D^{s}_{\nu\mu} (R^{\prime}).
\label{A2:16}
\eeq
When applied to the vacuum (\ref{A2:16}) gives  
\beq
U(R,0){A}^{\dagger}_{(m,s)}(\mathbf{p},\mu)\vert 0 \rangle  =
\sum_{\nu=-s}^s {A}^{\dagger}_{m,s}( R\mathbf{p},\nu)\vert 0 \rangle D^{s}_{\nu\mu} (R)
\label{A2:17}
\eeq
which means that either this vanishes or it transforms like a particle
of mass $m$, spin $s$, momentum $\mathbf{p}$ and magnetic quantum
number $\mu$.  The spin in these states created out of the vacuum is
the canonical spin.  This will vanish if there are no one-particle
intermediate states with mass $m$ and spin $s$ in (\ref{A2:1}).  While
the notation is purposely suggestive,
$A^{\dagger}_{m,s}(\mathbf{p},\mu)$ is {\it not a creation operator}.
In addition, $p^0$ is only equal to $\sqrt{\mathbf{p}^2+m^2}$ when
${A}^{\dagger}_{m,s}( \mathbf{p},\mu)$ is applied to the vacuum.

Because rotations do not change the time component, these operators
also satisfy
\[
[ \mathbf{P},{A}_{(m,s)}^{\dagger}(\mathbf{q},\mu)]   =
\int_{SU(2)} dR [\mathbf{P}, 
U(R,0){A}^{\dagger}(R^{-1}\mathbf{q} )U^{\dagger}(R,0)]
D^{s*}_{\mu s}(R) =
\]
\[
\int_{SU(2)} dR
U(R,0)[ R \mathbf{P},  {A}^{\dagger}(R^{-1}\mathbf{q} )]U^{\dagger}(R,0)
D^{s*}_{\mu s}(R) =
\]
\beq
 RR^{-1}\mathbf{q} \int_{SU(2)} dR
U(R,0)  {A}^{\dagger}(R^{-1}\mathbf{q} )U^{\dagger}(R,0)
D^{s*}_{\mu s}(R) =
\mathbf{q}
{A}^{\dagger}_{(m,s)} (\mathbf{q},\mu).
\label{A2:18}
\eeq
The normalization can be chosen so the states created out of the
vacuum have the normalization (\ref{S1:8}). 

The construction above cannot be used in the ``light-front'' case due to the
integral over $p^-$ in (\ref{A2:9}). However in that case it is enough
to project out the magnetic quantum number using a rotation
$R_z(\phi)$ which leaves $p^-$ in (\ref{A2:9}) unchanged.  The spin
is identified with the highest non-zero weight, $s=\mu_{max}$. In this case
equation(\ref{A2:13}) is replaced by
\beq
\tilde{{A}}^{\dagger}_{m,s}( \tilde{\mathbf{p}},\mu) := \int_{0}^{2\pi}
U(R_z(\phi),0)\tilde{{A}}^{\dagger}(R_z(\phi)^{-1}\mathbf{p} )U^{\dagger}(R_z(\phi),0)
e^{-i \mu \phi}
{d\phi \over 2 \pi},
\label{A2:19}
\eeq
(\ref{A2:17}) and (\ref{A2:18}) are replaced by
\beq
[ \tilde{\mathbf{P}},\tilde{{A}}_{(m,s)}^{\dagger}(\tilde{\mathbf{q}},\mu)]   =
\tilde{\mathbf{q}}
\tilde{{A}}^{\dagger}_{(m,s)} (\tilde{\mathbf{q}},\mu),
\label{A2:20}
\eeq
\beq
U(\Lambda_K,0) \tilde{{A}}^{\dagger}_{(m,s)} (\tilde{\mathbf{q}},\mu)
U^{\dagger}(\Lambda_K,0) = \tilde{{A}}^{\dagger}_{(m,s)} (\tilde{\Lambda}_k q,\mu),
\label{A2:21}
\eeq
and
\beq
U(R_z(\phi),0) \tilde{{A}}^{\dagger}_{(m,s)}
(\tilde{\mathbf{q}},\mu) U^{\dagger} (R_z(\phi),0) =
\tilde{{A}}^{\dagger}_{(m,s)} (R_z(\phi)\tilde{\mathbf{q}},\mu)
e^{i\mu\phi}.
\label{A2:22}
\eeq
The proof (\ref{A2:20}) is essentially the same as the proof of (\ref{A2:18}).
To show (\ref{A2:22}) note that for rotations about the $z$ axis 
\beq
B_F(p/m) R_z = R_z B_F (R_z^{-1}p/m)
\label{A2:23}
\eeq
where $B_F(p/m)$ is a light-front boost.  It follows that
\[
U(B_F(p/m),0)\tilde{{A}}^{\dagger}_{m,s}( \tilde{\mathbf{q}},\mu)
U(B_F(p/m),0)^{\dagger} =
\]
\[
\int_{0}^{2\pi}
U(B_F(p/m) R_z(\phi),0)\tilde{{A}}^{\dagger}(R_z(\phi)^{-1}\mathbf{q} )U^{\dagger}(
B_F(p/m)R_z(\phi),0)
e^{-i \mu \phi} =
\]
\[
\int_{0}^{2\pi}
U(R_z(\phi) B_F(R_z^{-1}(\phi)p/m),0)\tilde{{A}}^{\dagger}(R_z(\phi)^{-1}\mathbf{q} )U^{\dagger}(R_z(\phi) B_F(R_z^{-1}(\phi)p/m),0)
e^{-i \mu \phi}=
\]
\[
\int_{0}^{2\pi}
U(R_z(\phi),0)U( B_F(R_z^{-1}(\phi)p/m),0)\tilde{{A}}^{\dagger}(R_z(\phi)^{-1}\mathbf{q} )U(B_F(R_z^{-1}(\phi)p/m),0))^{\dagger}
U^{\dagger}(R_z(\phi),0)^{\dagger} 
e^{-i \mu \phi}=
\]
\[
\int_{0}^{2\pi}
U(R_z(\phi),0)U(,0)\tilde{{A}}^{\dagger}( B_F(R_z^{-1}(\phi)p/m)R_z(\phi)^{-1}\mathbf{q} )
U^{\dagger}(R_z(\phi),0)^{\dagger} 
e^{-i \mu \phi}=
\]
\[
\int_{0}^{2\pi}
U(R_z(\phi),0)U(,0)\tilde{{A}}^{\dagger}( R_z(\phi)^{-1} B_F(p/m)\mathbf{q} )
U^{\dagger}(R_z(\phi),0)^{\dagger} 
e^{-i \mu \phi}=
\]
\beq
\tilde{{A}}^{\dagger}_{m,s}( \tilde{\mathbf{
B}}_F(p/m)q ,\mu)
\label{A2:24}
\eeq
which proves (\ref{A2:21}). For (\ref{A2:22}) note that 
\[
U(R_z(\phi'),0) \tilde{{A}}^{\dagger}_{(m,s)}
(\tilde{\mathbf{q}},\mu) U^{\dagger} (R_z(\phi'),0) =
\]
\beq
\int_{0}^{2\pi}
U(R_z(\phi+\phi'),0)\tilde{{A}}^{\dagger}(R_z(\phi)^{-1}\mathbf{p} )U^{\dagger}(R_z(\phi+\phi'),0)
e^{-i \mu \phi}
{d\phi \over 2 \pi}.
\label{A2:25}
\eeq
Let $ \phi''= \phi + \phi'$ so (\ref{A2:25})) becomes 
\[
\int_{0}^{2\pi}
U(R_z(\phi''),0)\tilde{{A}}^{\dagger}(R_z(\phi''-\phi')^{-1}\mathbf{p} )U^{\dagger}(R_z(\phi''),0)
e^{-i \mu (\phi''-\phi')}
{d\phi'' \over 2 \pi}
\]
\[
\int_{0}^{2\pi}
U(R_z(\phi''),0)\tilde{{A}}^{\dagger}(R_z(\phi'')^{-1}R_z(\phi')\mathbf{p} )U^{\dagger}(R_z(\phi''),0)
e^{-i \mu (\phi''}e^{i\phi' \mu)} =
\]
\beq
\tilde{{A}}^{\dagger}_{(m,s)} (R_z(\phi')\tilde{\mathbf{q}},\mu)
e^{i\mu\phi'}.
\label{A2:26}
\eeq
The normalization can be chosen so these states created out of the
vacuum have the normalization (\ref{S1:8}).

\bibliography{collins.bib}
\end{document}